\documentclass[manuscript]{aastex}

\usepackage[english]{babel} 
\usepackage{longtable}
\usepackage{color}
\usepackage{ulem}

\shorttitle{Time evolution of a viscous protoplanetary disk with a free geometry: toward a more self-consistent picture
}
\shortauthors{Bailli\'e and Charnoz}
\begin{document}


\title{Time evolution of a viscous protoplanetary disk with a free geometry: toward a more self-consistent picture
}


\author{K\'evin Bailli\'e and S\'ebastien Charnoz}
\affil{Laboratoire AIM-LADP, Universit\'e Paris Diderot/CEA/CNRS, 91191 Gif sur Yvette, France.}
\email{kevin.baillie@cea.fr}



\begin{abstract}
Observations of protoplanetary disks show that some characteristics seem recurrent, even in star formation regions that are physically distant such as surface mass density profiles varying as $r^{-1}$, or aspect ratios about 0.03 to 0.23. Accretion rates are also recurrently found around $10^{-8} - 10^{-6}~\mathrm{M_{\odot}~yr^{-1}}$ for disks already evolved (\cite{isella09,andrews09,andrews10}). Several models have been developed in order to recover these properties. However, most of them usually simplify the disk geometry if not its mid-plane temperature. This has major consequences for modeling the disk evolution over million years and consequently planet migration. In the present paper, we develop a viscous evolution hydrodynamical numerical code that determines simultaneously the disk photosphere geometry and the mid-plane temperature. We then compare our results of long-term simulations with similar simulations of disks with a constrained geometry along the \cite{chiang97} prescription (dlnH/dlnr = 9/7). We find that the constrained geometry models provide a good approximation of the disk surface density evolution. However, they differ significantly regarding the temperature time evolution. In addition, we find that shadowed regions naturally appear at the transition between viscously dominated and radiation dominated regions that falls in the region of planetary formation. We show that $\chi$ (photosphere height to pressure scale height ratio) cannot be considered as a constant, consistently with \cite{watanabe08}. Comparisons with observations show that all disk naturally evolve toward a shallow surface density disk ($\Sigma \propto r^{-1}$). The mass flux across the disk stabilizes in about 1 million year typically.

\end{abstract}


\section{Introduction} \label{intro}
Observations of gas-rich circumstellar disks enable to constrain the outer regions of these disks whereas the inner regions remain more cryptic. In particular, the latest observations of the Taurus \citep{isella09} and Ophiuchus \citep{andrews09,andrews10} young stars provided details about the large scale morphology of protoplanetary disks: surface mass densities, temperatures, photosphere heights, accretion rates... Among the recurrent characteristics, \cite{mundy00} and \cite{garaud07} report shallower surface mass density profiles than the usual Minimum Mass Solar Nebula \citep{weiden77, hayashi81}, while measured flaring angles of 0.03 to 0.23 (and up to 0.26 from the \cite{lagage06} VISIR observation of HD97048) may provide additional constraints for the photosphere height of the disks. The accretion rates are also recurrently found around $10^{-8} - 10^{-6}~\mathrm{M_{\odot}~yr^{-1}}$ providing upper values on the age of the disks.

The purpose of the present paper is to establish the importance of a realistic protoplanetary disk geometry in order to calculate its thermodynamics and dynamical evolution. Its temperature will actually govern how the disk will spread and therefore its mass distribution. We will thus detail a numerical model for the dynamical and thermodynamical evolution of a disk around a classical T Tauri type star over timescales about the same order of magnitude as the disk lifetime, while taking into account the coupling between the disk photosphere geometry and its temperature profile.

Numerous previous studies tried to approach the problem using different approximations. For example, some studies neglect some of the viscous effects: \cite{dullemond01}, \cite{jc04} and \cite{jc08} modeled passive disks while \cite{dalessio98}, \cite{hughes10} and \cite{bitsch12} neglected only the viscous spreading and kept the viscous heating contribution. Other very constraining hypotheses set surface mass density profiles \citep{calvet91} or mid-plane temperature profiles \citep{hughes10}. Neglecting the disk irradiation by the star simplifies also the problem as the viscous heating only depends on the surface mass density and the viscosity, and not on the disk shape. However, this approximation is only valid in the case of a dominating viscous heating \citep{hueso05}. Actually, most observational constraints on the physical properties of the disks are provided by the study of the outer regions, where the viscous heating is negligible compared to the irradiation heating. It is therefore necessary to consider both the viscous heating and the irradiation heating. Numerous studies impose a uniform and constant grazing angle (as in \cite{ciesla09} and \cite{zhu08} using the 0.05 rad value derived by \cite{brauer08}) or assume a photosphere height profile in $r^{9/7}$ \citep{hueso05,birnstiel10} as derived by \cite{chiang97} in the case of a steady state with a surface mass density in $r^{-3/2}$. While these models provide a good knowledge of the outermost regions having reached their steady state, the viscous evolution timescales suggest that such a steady state is maintained only for a short period of time before the disk gets photo-evaporated in a few million years. It is then necessary to focus on the transitory evolution of the disk before it reaches the steady state since it appears that the planet cores will accrete very quickly in the early evolution of the disk. The numerical code we detail in this paper does not rely on any steady state analytical equation and sets free a number of parameters such as the geometric structure by coupling it to the thermodynamical structure on one hand and coupling the thermodynamical evolution to the dynamical evolution on the other hand. Comparisons of the obtained steady state asymptotic behavior with analytical developments and actual observations will provide validation of that code in order to further study the formation of the first solids in future papers: \cite{hasegawa112} showed for example that planetary traps can be generated from irregularities in temperature or density radial distributions.

In the present paper, we calculate the disk photosphere and pressure scale heights jointly with its mid-plane temperature at every time step. We can therefore study the transitions between zones dominated by viscous heating and stellar irradiation. Using pre-existing semi-analytical models, we have built a hydrodynamical evolution code that properly includes the disk geometry and eliminates some assumptions and fixed parameters, allowing to derive self-consistently the disk structure. The dynamical and thermodynamical parameters are coupled through the turbulent viscosity that drives the viscous heating and the viscous spreading. We will confront the observational data with our numerical models and show the convergence toward a steady state with a surface-mass density decreasing as $r^{-1}$ independently from the initial density profile. Of course, the present model makes numerous approximations of some of the disk aspects but a special effort has been made for a consistent coupling of the dynamical and thermodynamical evolutions.

We present the physical model and the numerical code in Section \ref{methods}. We apply our numerical evolution to a standard protoplanetary disk model, the Minimum Mass Solar Nebula (MMSN) from \cite{weiden77} around a typical T Tauri type star in Section \ref{mmsn} and the sensitivity to the initial conditions in Section \ref{ini}. We discuss the importance of calculating self-consistently the geometric structure of the disk in Section \ref{geo}. We then compare our simulated disks to analytical asymptotic solutions on one hand and observations from \cite{isella09}, \cite{andrews09} and \cite{andrews10} on the other before discussing the consequences on the disk properties in Section \ref{obs}.

\section{Methods} \label{methods}

\subsection{Viscous $\alpha$ disk}
We consider the protoplanetary disks to be turbulent and to follow an "$\alpha$" prescription as defined in \cite{shakura73}. At a given distance $r$ from the star (surface effective temperature $T_{*}$, mass $M_{*}$, radius $R_{*}$ and luminosity $\mathcal{L}_{*}$), the viscosity $\nu(r)$ is then defined as:\\
\begin{equation}
\label{eqnunu}
\nu(r) ~= ~\alpha_{\mathrm{visc}} ~c_{s}(r) ~h_{pr}(r),
\end{equation}\\
where $c_{s}(r)$, the local isothermal sound speed, is the characteristic turbulent velocity and is defined as $\sqrt{\frac{k_{B}T_{m}(r)}{\mu m_{p}}}$, with $k_{B}$ the Boltzmann constant, $T_{m}(r)$ the mid-plane temperature at the distance $r$, and $\mu$ the mean molecular weight ($\mu = 2.3$ for a fully molecular gas of cosmic composition) in units of the proton mass $m_{p}$; and $h_{pr}(r)$ is the local pressure scale height of the disk (characteristic mixing length): $h_{pr}(r) ~= \frac{c_{s}(r)}{\Omega(r)}$, where $\Omega(r)$ is the Keplerian angular velocity $\sqrt{\frac{GM_{*}}{r^{3}}}$. 
The amount of turbulence is controlled by the free parameter $\alpha_{\mathrm{visc}}$ which was found around $10^{-2}$ for T Tauri stars by \cite{hartmann98}. Though this parameter remains fairly unconstrained, magneto-hydrodynamical numerical simulations from \cite{fromang06} showed that $\alpha_{\mathrm{visc}}$ may be in the range 0.001 to 0.01. Therefore, we take $10^{-2}$ as a default value for our disks, and for the purpose of this paper we assume $\alpha_{\mathrm{visc}}$ to be spatially uniform and temporally constant as we do not treat the case of disks with dead-zones.

\subsection{Temporal evolution}
Previous studies (Section \ref{intro}) modeled protoplanetary disks with strong dynamical or thermodynamical constraints (imposed accretion rates, temperature or surface mass density profiles). Cumulating various hypotheses may result in losing the model consistency. In this paper, we suggest freeing most of these constraints by coupling the dynamical and thermodynamical evolutions thanks to the viscosity that drives both the disk spreading and its temperature. Given a certain initial surface mass density profile $\Sigma(r)$, we derive the mid-plane temperature for any distance to the star (using a method detailed in Section \ref{temp}), and compute the local viscosities. The disk viscous spreading is computed following the \cite{lyndenbellpringle74} equation:
\begin{equation}
\frac{\partial \Sigma(r,t)}{\partial t} = \frac{3}{r}\frac{\partial}{\partial r}\left(\sqrt{r} \frac{\partial}{\partial r} \left( \nu(r,t) \Sigma(r,t) \sqrt{r}\right) \right)
\label{lb74}
\end{equation}

It follows that the local mass flux can be expressed as:
\begin{equation}
\label{eqflux}
F_{v}(r) = -6 \pi \sqrt{r} \frac{\partial}{\partial r} \left( \nu(r) \Sigma(r) \sqrt{r} \right)
\end{equation}

Our numerical code consists in applying Equation \ref{lb74} to a 1-dimension grid of masses logarithmically distributed in radius between 0.01 AU (or $R_{*}$, whichever is greater) and 1000 AU. As we do not want the inner boundary to steal mass from the star, we impose that the flux at the inner edge cannot be directed outward. However, in order to remove the bias due to these inner boundary conditions, we only visualize our disks starting at 0.1 AU. The mass flux at the innermost location provides the mass accretion rate of the disk.

\subsection{Disk geometry and temperature}\label{temp}
We consider that the temperature in the mid-plane, $T_{m}(r)$, results from the combination of viscous heating, stellar irradiation heating and radiative cooling in the mid-plane.

\begin{figure}[htbp!]
\center
\includegraphics[width=16cm]{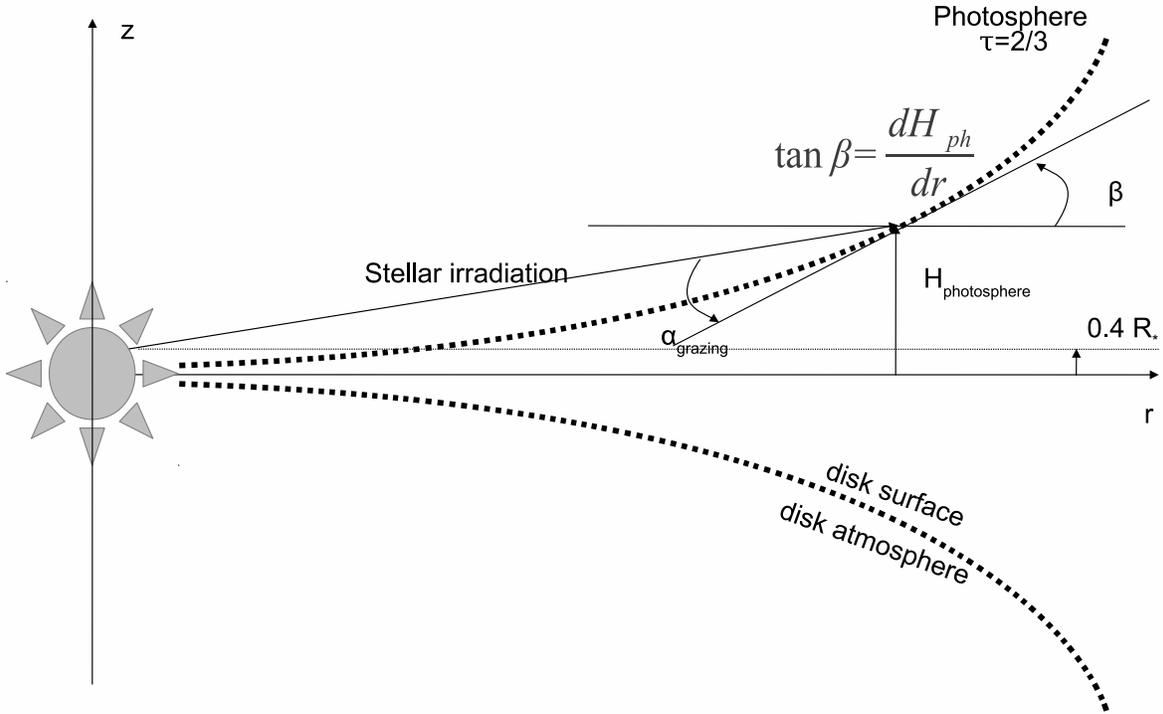}
\caption{Geometric model of an irradiated protoplanetary disk. After integration over the star surface of emission, we consider the radiation emitted at $z = 0.4 R_{*}$.}
\label{figalphagr}
\end{figure}

Rather than using the prescription from \cite{chiang97} that assumes that the photosphere height follows a power-law profile in $r^{9/7}$, we define for each radius $r$ the angle at which the star sees the photosphere as the grazing angle $\alpha_{gr}(r)$. Comparisons between an imposed geometry following that prescription and a free geometry calculated along with a consistent temperature are shown in Section \ref{geo}, along with a discussion of the necessity of these geometric refinements. Figure \ref{figalphagr} shows the geometric definition of the grazing angle and how it is related to the photosphere height $H_{ph}(r)$: we can therefore derive Equation \ref{alphagr}:

\begin{equation}
\alpha_{gr}(r) = \arctan\left(\frac{dH_{ph}}{dr}(r)\right) - \arctan\left(\frac{H_{ph}(r)-0.4 R_{*}}{r}\right)
\label{alphagr}
\end{equation}

This angle actually governs the amount of energy provided to the disk by the star. At a given location, a positive grazing angle would result in the disk being irradiated. Regions not irradiated could be shadowed by inner regions.

\begin{figure}[htbp!]
\center
\includegraphics[width=18cm]{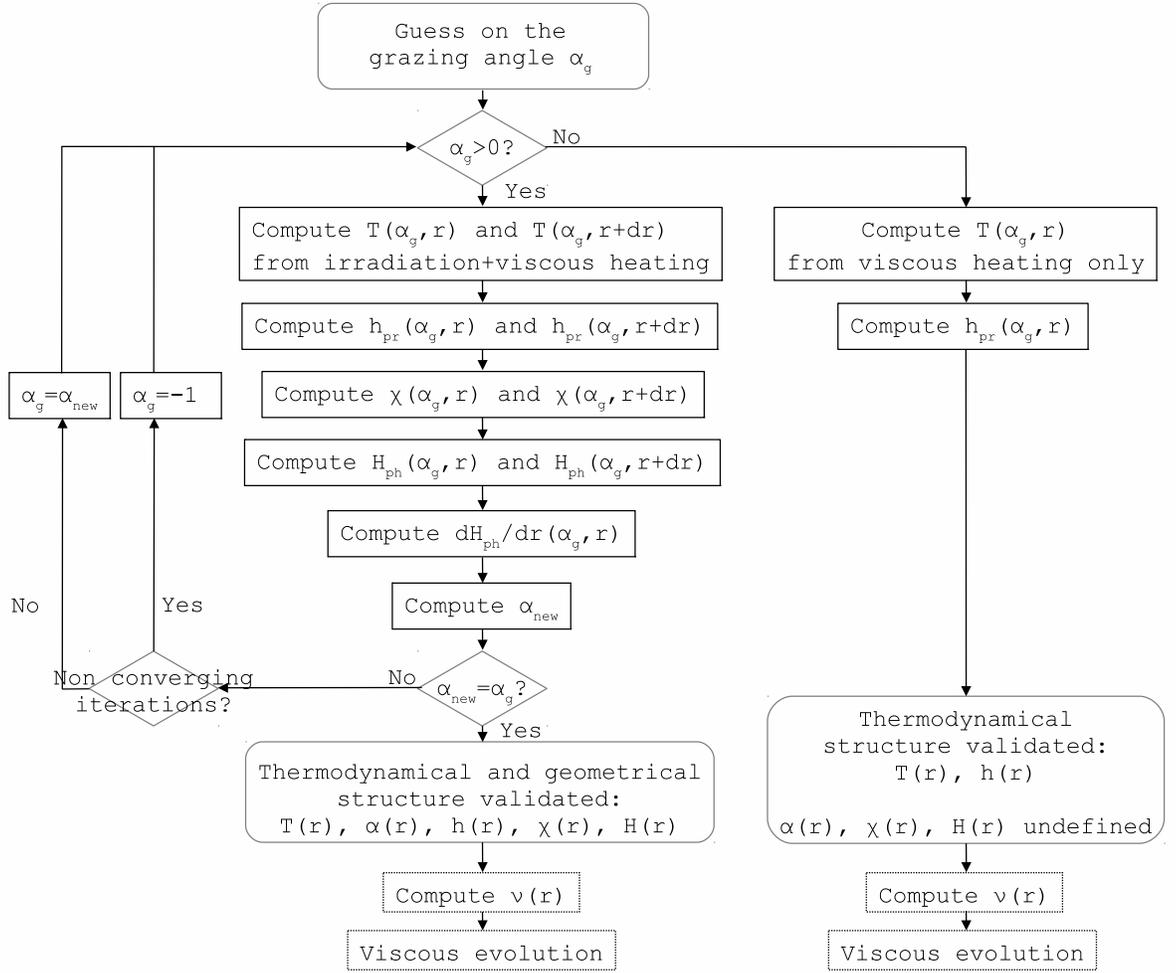}
\caption{Algorithm description for the determination of the joint structures in geometry and temperature.}
\label{flowchart}
\end{figure}

We then use Equation 18 from \cite{calvet91} to calculate the temperature in the mid-plane: this model considers the heating/cooling contributions that we have listed as contributors to the mid-plane temperature. We are therefore able to recalculate the viscous heat flux self-consistently from the surface mass density obtained after temporal evolution from \cite{lyndenbellpringle74} equation:
\begin{equation}
F_{v}(r) = \frac{1}{2} \Sigma(r) \nu(r) \left( R \frac{\mathrm{d}\Omega}{\mathrm{d}r} \right)^{2} = \frac{9}{4} \Sigma(r) \nu(r) \Omega^{2}(r)
\end{equation}
As this viscous contribution term depends on the mid-plane temperature itself through the viscosity (see Equation \ref{eqnunu}), Equation 18 from \cite{calvet91} becomes an implicit equation on the mid-plane temperature. The opacities are estimated for a typical gas to dust mass ratio of $1\%$ and for a wavelength corresponding to a temperature of $4000$ K for the star emission and $300$ K for the disk re-radiation. In addition, this temperature is highly affected by the geometry of the disk as the grazing angle controls the efficiency of the irradiation heating. Therefore, the geometrical structure (photosphere and pressure heights) is determined jointly with the temperature by iterating numerically on the grazing angle value. The algorithm is described in Figure \ref{flowchart}. 

From an initial guess on the grazing angle at a distance $r$, we can solve the implicit equation providing the corresponding mid-plane temperature and pressure scale height. Considering a hydrostatic equilibrium, the vertical density distribution follows a Gaussian, and we can use Equation A9 from \cite{dullemond01} to calculate the ratio $\chi$ of the photosphere height to the pressure scale height:

\begin{equation}
\label{chi}
1- \mathrm{erf}\left(\frac{\chi(r)}{\sqrt{2}}\right) = \frac{2\alpha_{gr}(r)}{\Sigma(r) \kappa_{P} (T_{*})},
\end{equation}

where $\kappa_{P} (T_{*})$ is the Planck mean opacity at stellar temperature $T_{*}$. It is assumed here that the disk vertical density profile is the same as an isothermal vertical structure at the mid-plane temperature. This is of course an approximation which is reasonable below a few pressure scale heights, where most of the disk mass is located.

We can then estimate the corresponding presumed photosphere height $H_{ph}$ at each radial location and therefore access $\frac{dH_{ph}}{dr}$. Applying Equation \ref{alphagr}, we can verify whether the presumed grazing angle has the required precision or if we should iterate on it. The impossibility to solve that problem for any positive value of the grazing angle results in a disk column that is not directly irradiated by the star, and therefore we remove the irradiation heating term from the mid-plane temperature equation. The disk structure is then considered in its whole: geometric shape, vertical thickness, mid-plane temperature and viscosity, all related to provide a consistent geometrical-thermodynamical structure, resulting in a significant improvement compared to most of the previous studies that required to fix at least one of these quantities. However, taking into account the possible shadowing effects of more interior regions results in numerical instabilities which can be solved by increasing the radial resolution despite an explosive computational time. Therefore, we do not consider such geometrical refinements in the scope of this paper and leave them for a further study.

\subsection{Sublimation zone}
Though we assume the dust to have a spatially uniform opacity, the physical composition of the disk can vary drastically when the temperature reaches the dust sublimation temperature, considered here at 1500 K (reasonable for the most refractory elements), implying a drop in opacity. Therefore, the mid-plane temperature must be recalculated in these regions, taking into account that the disk is slightly less opaque than initially estimated. We make the assumption that the opacity is mainly due to the dust in the regions where $T_{m}~<~1500~K$. When the temperature is higher, we first consider the new theoretical temperature that would result from the opacity being entirely due to the gas (taken here to be 1\% of the dust opacity after \cite{dalessio01}). We consider the dust to be totally or partially sublimated based on the new temperature being greater than the sublimation temperature or lower. In the later case, we consider the mid-plane temperature as equal to the sublimation temperature and derive the proportion of dust that is sublimated by estimating the required opacity to obtain the sublimation temperature. Though we do not reach the level of refinement of \cite{ruden91} in the local estimation of the opacities, our process still allows to take into consideration the most important variation of the opacities. Further work on other opacity transitions will be developed in a next paper.

In the early ages, we expect the mid-plane temperature to pass 1500 K in the innermost regions and therefore we expect a sublimation zone to exist, whereas it is likely that this zone will disappear as the disk cools down in a few 100,000 years.

\section{Standard nebula evolution} \label{mmsn}

We use the "Minimum Mass Solar Nebula" (MMSN) model of the Solar proto-system described in \cite{weiden77} and \cite{hayashi81} as a fiducial case. The MMSN surface mass density is given by:
\begin{equation}
\label{eqmmsn}
\Sigma (r) = 17,000 \left(\frac{r}{1 \mathrm{AU}}\right)^{-3/2} \mathrm{kg\cdot m^{-2}}
\end{equation}

The MMSN is shown by \cite{vorobyov07} to be consistent with an intermediate stage of a protoplanetary disk between a collapsing molecular cloud and a steady state disk, under self-regulated gravitational accretion. We use the MMSN around a classical T Tauri type young star ($M_{*} = 1 M_{\odot}$, $R_{*} = 3 R_{\odot}$, $T_{*} = 4000 \mathrm{K}$ and $\mathcal{L}_{*} = 4 \pi R_{*}^{2} \sigma_{B} T_{*}^{4}$) for the initial condition. We do not consider the gravitational collapse of the molecular cloud in the scope of this paper since models of \cite{vorobyov07} for example show that there is a stage in collapse evolution for which the surface mass density follows a power-law in $r^{-1.5}$. In addition, for a fixed $\alpha_{\mathrm{visc}}$, the various initial conditions converge toward a power-law profile as will be shown in Section \ref{ini}. Therefore, the MMSN profile makes an as good initial profile as any other snapshot that could have been taken in the disk evolution. In addition, this initial configuration makes sense in order to better compare our results with previous studies. The initial state is displayed in black in Figures \ref{mmsnsigma}-\ref{mmsnflux} and \ref{mmsntemp}-\ref{mmsnhphoto}.

\subsection{Surface mass density evolution: toward a steady state $\Sigma \propto r^{-1}$}
\begin{figure}[htbp!]
\center
\includegraphics[width=16cm]{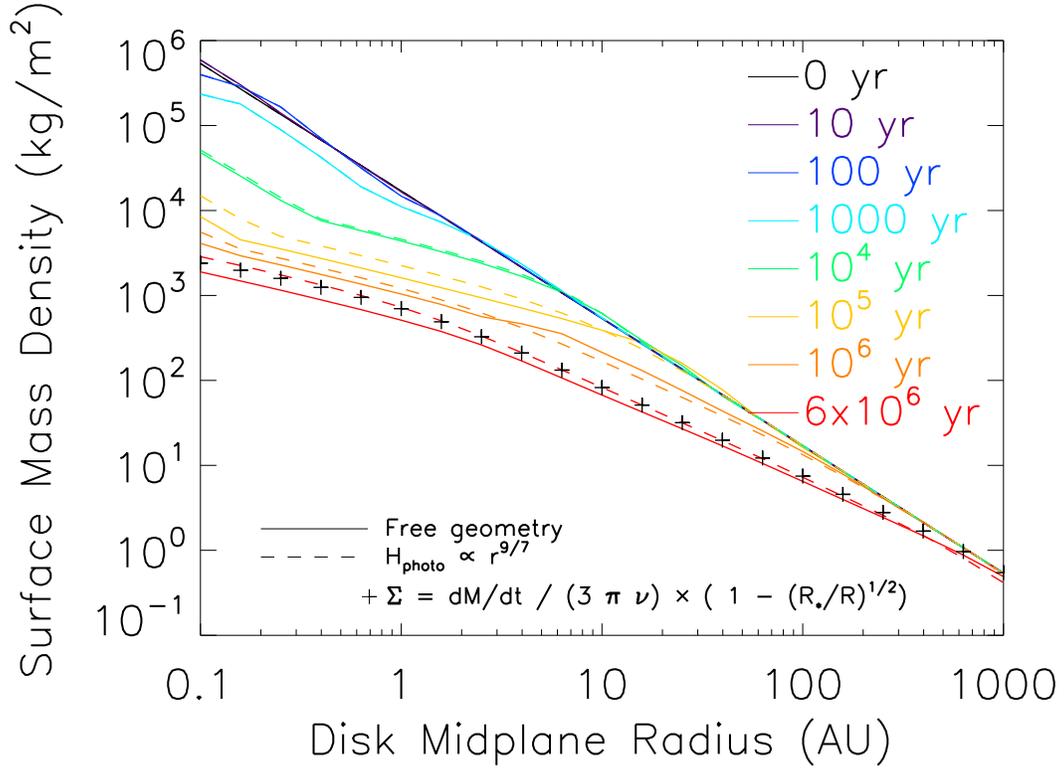}
\caption{Surface mass density profile evolution for a Minimum Mass Solar Nebula for a self-consistently calculated geometry (solid line) or an imposed geometry following $r^{9/7}$ (dashed line). The '+' show the surface mass density profile derived from the conservation of the angular momentum (\cite{lyndenbellpringle74}).}
\label{mmsnsigma}
\end{figure}

The evolution of the surface mass density profiles over 6 million years is presented in Figure \ref{mmsnsigma}. We observe that the profile gets shallower and tends to a power-law $\Sigma (r) \propto r^{-1.03}$ as reported by \cite{mundy00} and \cite{garaud07}. Despite this timescale is somewhat longer than a disk characteristic lifetime (\cite{font04}, \cite{alexander07, alexander09} and \cite{owen10} showed that photo-evaporation will dissipate the disk in a few million years), our numerical simulations are kept running until the disk reaches a steady state. We notice that the profile gets shallower in the inner regions at the beginning of its evolution, while keeping the memory of the initial shape in the outer regions, as we can see from the break in the power-law in the surface mass density profile. It takes less than 100,000 years to reach a power-law index of $-1$ in the region between 10 and 100 AU. This break drifts outward until it reaches the outer maximal radius of our simulations at 1000 AU.

Figure \ref{mmsnsigma} also shows the difference in evolution between a \cite{chiang97} geometry and a free geometry (i.e. self-consistently calculated): the surface mass density profiles can be well approximated by the imposed geometry on a large scale. In addition, the expression of the density profile $\Sigma = \frac{\dot{M}}{3 \pi \nu} \left( 1-\sqrt{\frac{R_{*}}{R}}\right)$ derived from the conservation of the angular momentum by \cite{lyndenbellpringle74} is also valid in first approximation.

\subsection{Mass-accretion rate evolution: toward a constant accretion rate}
\label{sectionmmsnflux}
\begin{figure}[htbp!]
\begin{center}$
\begin{array}{cc}
\includegraphics[width=8cm]{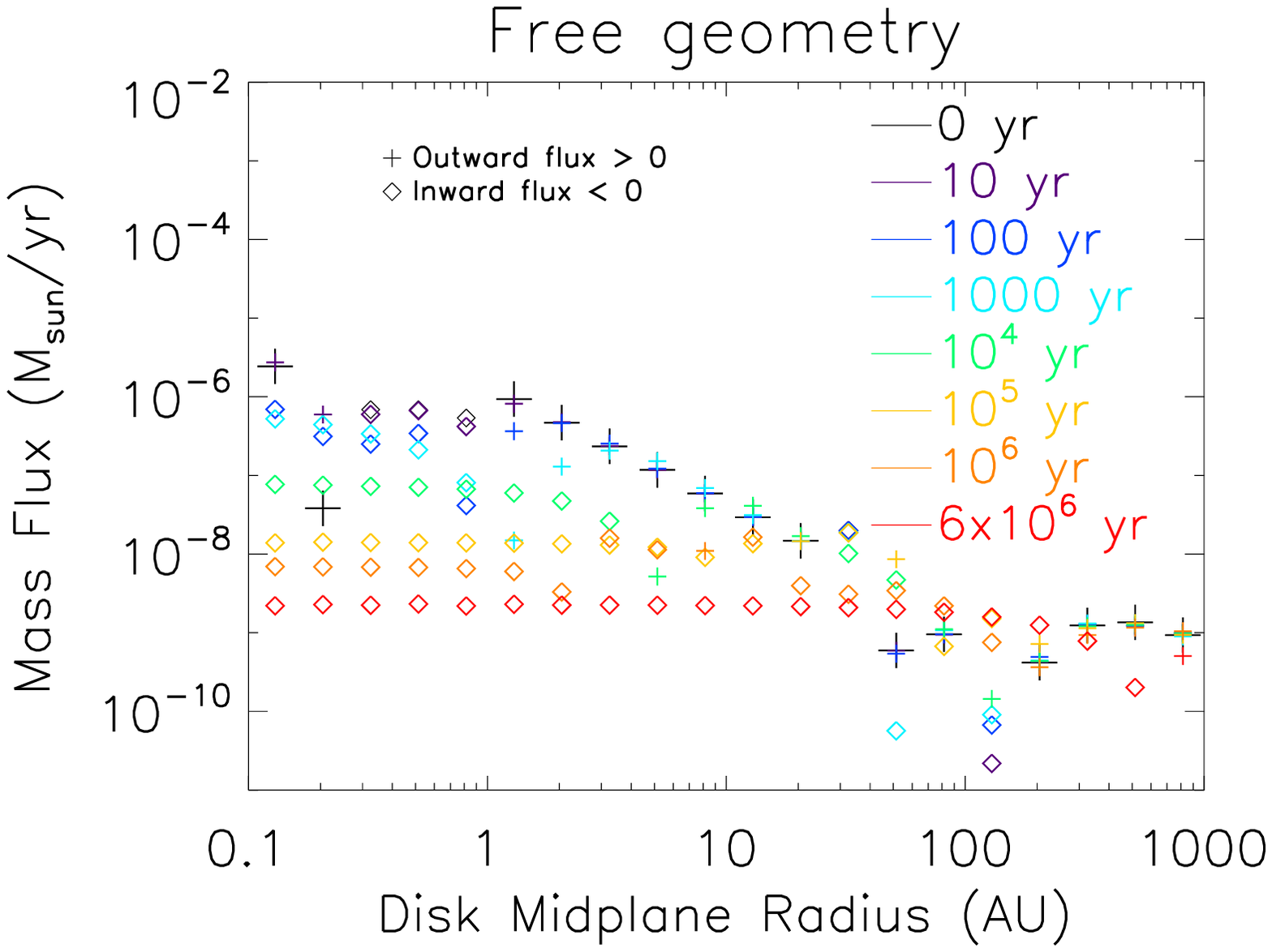}&
\includegraphics[width=8cm]{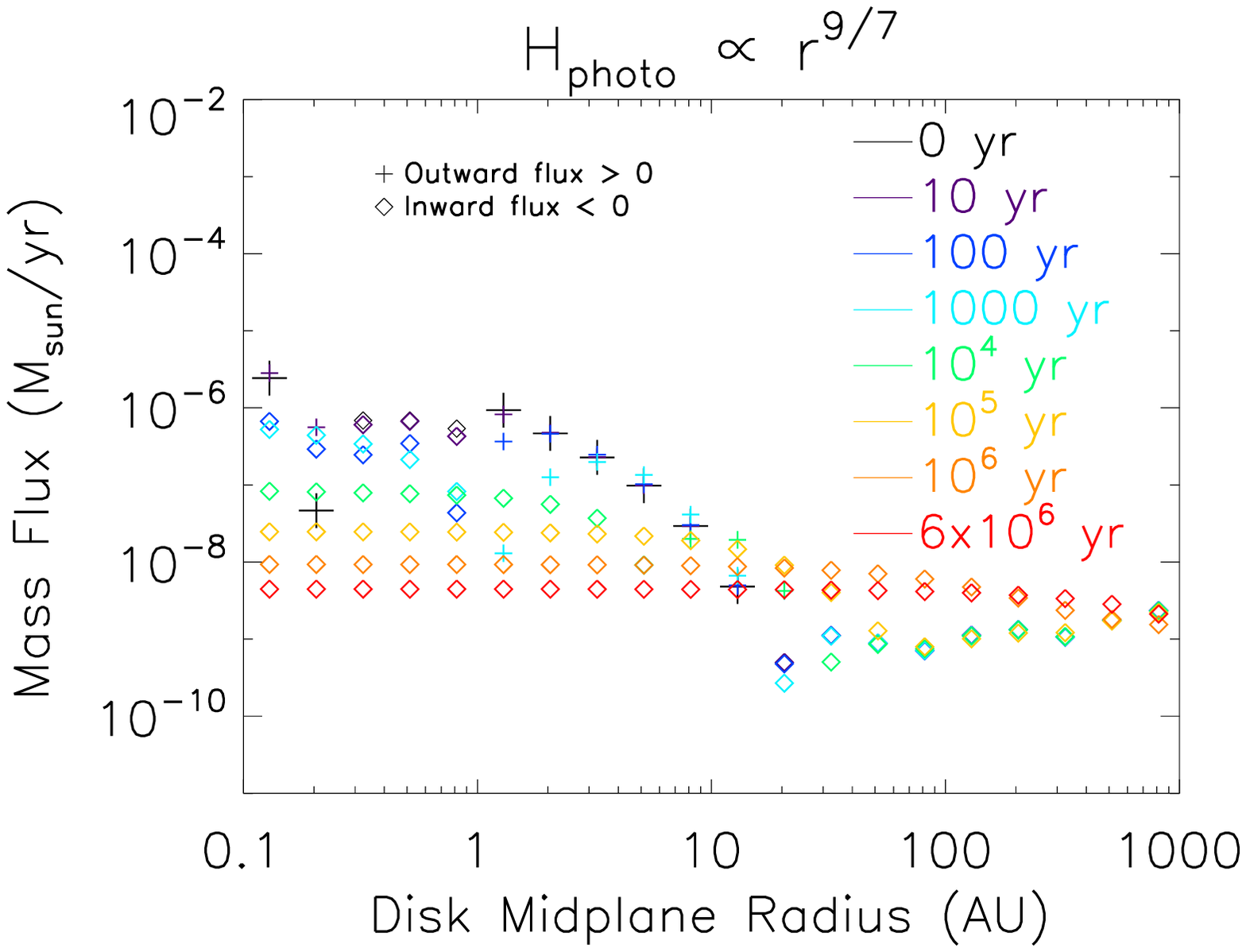}\\
\end{array}$
\caption{Mass flux profile evolution for a Minimum Mass Solar Nebula in the case of a self-consistently calculated geometry (left panel) and an imposed geometry following $r^{9/7}$ (right panel).}
\label{mmsnflux}
\end{center}
\end{figure}

Figure \ref{mmsnflux} shows the evolution of radial profiles for the mass fluxes. The initial disk spreads outward with amplitudes fairly higher than the usual observed fluxes \citep{gullbring98}. Rapidly after, there is a frontier, interior to which the disk flows inward onto the star, whereas at the exterior of it, the disk flows in the direction of the increasing radii. This frontier, located around 1 AU after 1000 years of evolution, moves outward and reaches the disk outer edge after 6 million years. The fluxes decrease in amplitude down to a few $10^{-9}~ \mathrm{M_{\odot}\cdot yr^{-1}}$, getting closer to the observed values, and they tend to get uniform over the disk. After the 1 million years, the disk is almost fully accretional. This uniform flux profile is synonymous of steady state thanks to Equation \ref{lb74}. From Equation \ref{steady2} and the asymptote found for the surface mass density profile, we would expect to find a temperature profile close to $T_{m}(r) \propto r^{-1/2}$.

Figure \ref{mmsnflux} confirms similar evolution in the first 100,000 years for different geometries and also shows a longer delay in reaching the steady state beyond 100 AU in the case of a free geometry.

\begin{figure}[htbp!]
\center
\includegraphics[width=16cm]{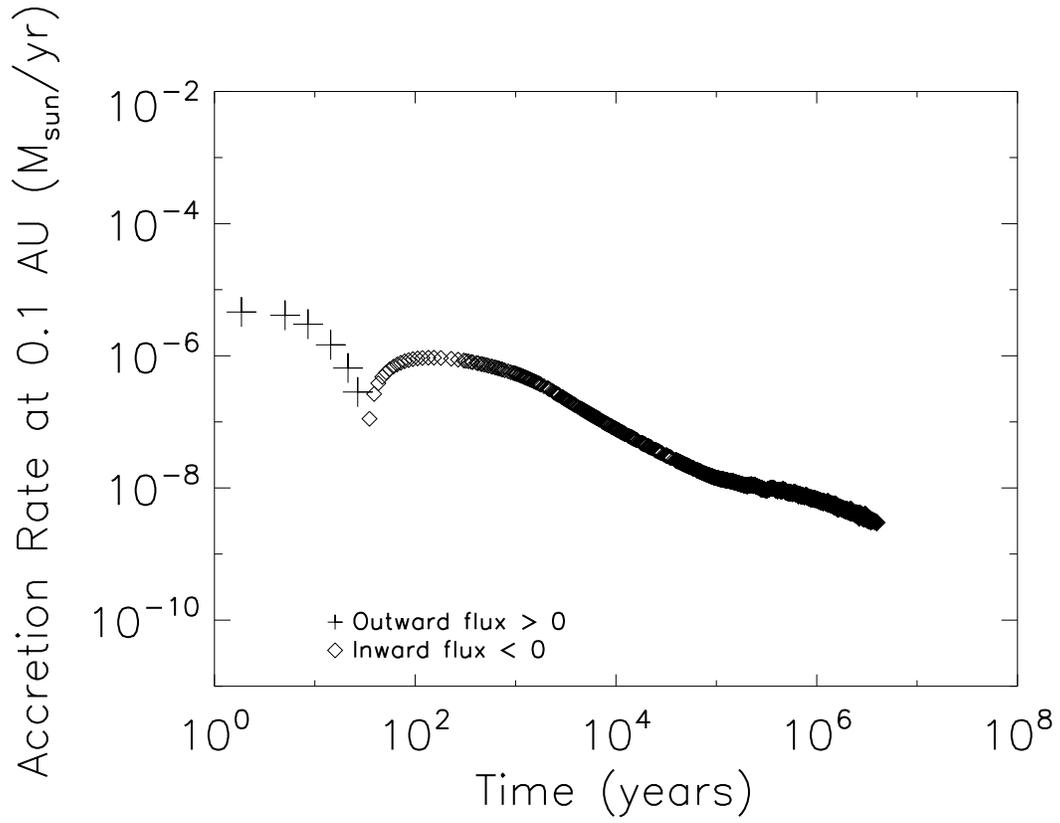}
\caption{Evolution of the mass flux at 0.1 AU for a Minimum Mass Solar Nebula.}
\label{mmsnfluxt}
\end{figure}

Considering the accretion rate at $0.1$ AU as a function of time (Figure \ref{mmsnfluxt}), we notice that the flux is directed outward for the first 60 years before turning inward and decreasing in amplitude as the disk material is falling on the star. This decreasing could be modeled using the following power-law $\dot{M}_{\mathrm{0.1~AU}}(t) \propto t^{-0.51}$, which would be consistent with the approximated solution of a one-dimension diffusion of the gas through our disk:
\begin{equation}
\dot{M}(r,t)  = \dot{M}_{0} \cdot \mathrm{erfc}\left(\frac{r}{2\sqrt{Dt}}\right) \sim \dot{M}_{0} \left(1-\frac{r}{\sqrt{D t \pi}}\right)
\end{equation}
with $\dot{M}_{0}$ a constant and $D$ the modeled diffusion coefficient.\label{mmsnfluxtemp}

This slope appears to be shallower than the estimation from \cite{hartmann98} whose values remain fairly uncertain partly due to the lack of precision on the birthline determination.

\subsection{Temperature structure: where the geometry matters}
\begin{figure}[htbp!]
\center
\includegraphics[width=16cm]{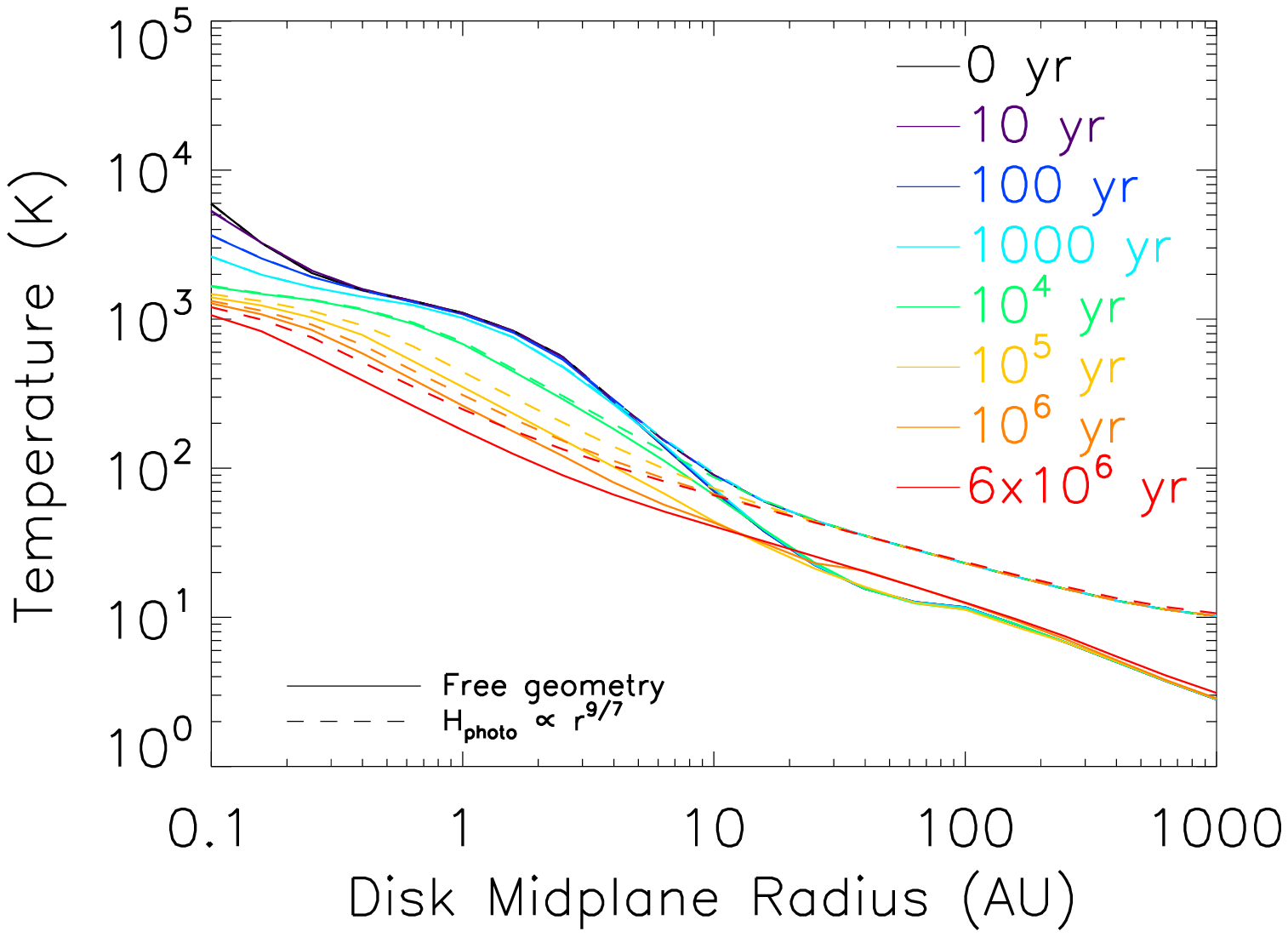}
\caption{Mid-plane temperature profile evolution for a Minimum Mass Solar Nebula in the case of a self-consistently calculated geometry and an imposed geometry following $r^{9/7}$.}
\label{mmsntemp}
\end{figure}

Figure \ref{mmsntemp} shows the evolution of the mid-plane temperature profile for a viscously evolving disk over 6 million years. This disk receives heat from the stellar irradiation and from the viscous heating.

\begin{figure}[htbp!]
\center
\includegraphics[width=16cm]{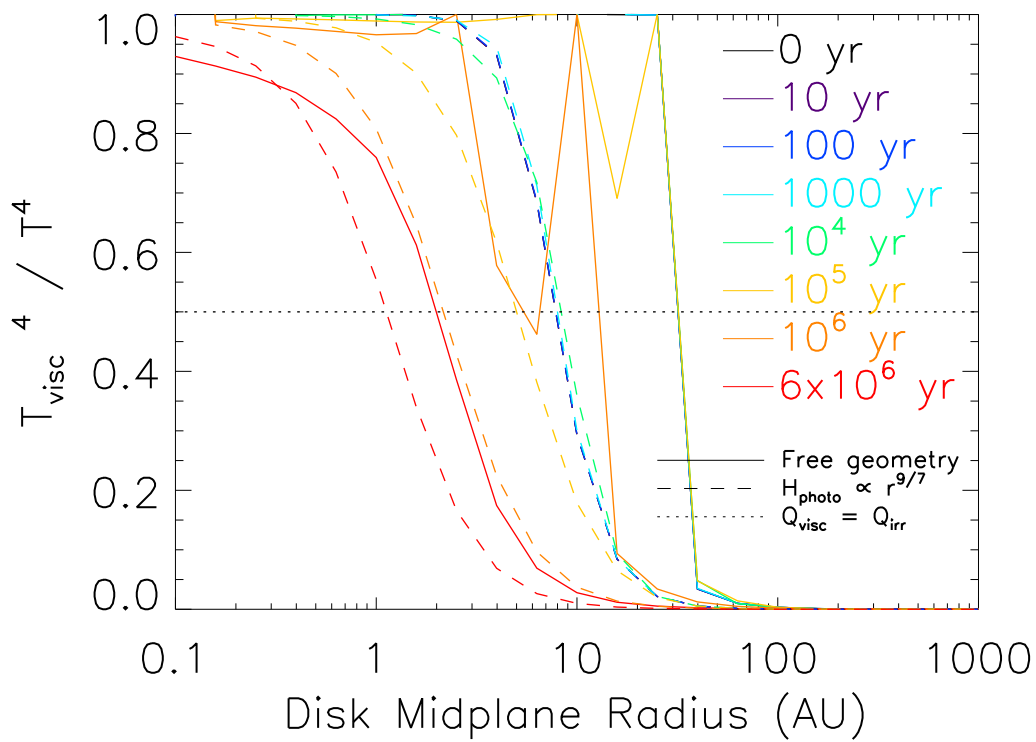}
\caption{Normalized contribution of the viscous heating to the mid-plane temperature. Dashed lines show the evolution of the viscous contribution for a imposed geometry (\cite{chiang97}) and (+) points show the evolution for a non-imposed and consistent geometry.}
\label{QTvm}
\end{figure}

In figure \ref{mmsntemp}, the initial temperature structure of the disk exhibits the different regions where each heating mode predominates: the outer regions are dominated by the irradiation heating, while the inner most regions are mainly dominated by the viscous heating. In the middle, both effects are competing. This is confirmed by Figure \ref{QTvm} showing the relative importance of the two heating processes in the disk for various times. We also notice the presence of a plateau in the initial temperature profile, located around the sublimation temperature, between 0.4 and 2 AU. This plateau reflects the effect of the sublimation of the dust grains on the vertical column. At this temperature, the opacity decreases and so does the efficiency of the viscous heating consequently. This results in creating a shallower plateau in the temperature distribution. The decrease of the surface mass density with time will affect the viscous heating primarily. Therefore, it is expected to observe a decrease in the steepness of the temperature profile with time. In the inner regions, the sublimation zone ($T \ge T_{sublimation} = 1500$ K) drifts inward until it passes the inner edge of our disk and it disappears completely after 100,000 years. The sublimation zone, on the other hand, extends initially up to 0.5 AU and is progressively narrowed until it disappears from the simulation when the inner edge temperature gets below 1500 K in a few million years. After 6 million years, in the case of a free geometry (Figure \ref{mmsntemp} - solid lines), the temperature profile asymptotic behavior with the radial distance can be modeled using a power-law $T_{m}(r)~\propto~ r^{-0.50}$, in agreement with the assumption that the steady state is reached (cf. Section \ref{mmsnfluxtemp}). This asymptotic behavior is consistent with the temperature profile for a flared outer disk, as estimated by \cite{kenyon87}: a constant $\alpha_{gr}$ yields to $T_{m}(r) \propto r^{-1/2}$.  The snowline (around 150 - 170 K, according to \cite{hayashi85}) is initially located around 6 AU and drifts inward to 1 AU in the steady state. Figure \ref{mmsntemp} allows to compare the temperature profiles for different geometries: an imposed geometry in $r^{9/7}$ exhibits a higher temperature in the outer regions and a slightly shallower asymptotic trend. This difference extends to a few tens of Kelvins in the region where irradiation heating and viscous heating are in competition. In the inner regions, the difference is more tenuous since the geometry impacts mainly the irradiation heating which is dominated by the viscous heating.

Figure \ref{QTvm} reflects the importance of viscous heating in the outer regions and irradiation heating in the inner ones. Between these regions, in the middle of the disk, there is a competition between these two contributions. That is also where shadowing phenomenons may occur as we can see from the peaks between 10 and 30 AU. The transition zone (intersection with the dotted line) where both contributions are of the same order of magnitude is located initially beyond 10 AU and moves progressively toward 2 AU after a few million years. This means that the entire planetary formation zone is dominated by viscous heating. This zone is a preferential place for generating temperature bumps that could create planetary traps for example. While an imposed geometry will clearly separate the zones of predominance for the two contributions (viscous heating inside and irradiation heating outside), the transition zone is less clearly delimited in the case of a self-consistently calculated geometry as shadowed regions may appear in there. Indeed, some regions appear naturally not irradiated, generating a brutal increase in the ratio of the heating processes.

\subsection{Pressure and photosphere heights: shadow regions affect the thermodynamics}
\begin{figure}[htbp!]
\center
\includegraphics[width=16cm]{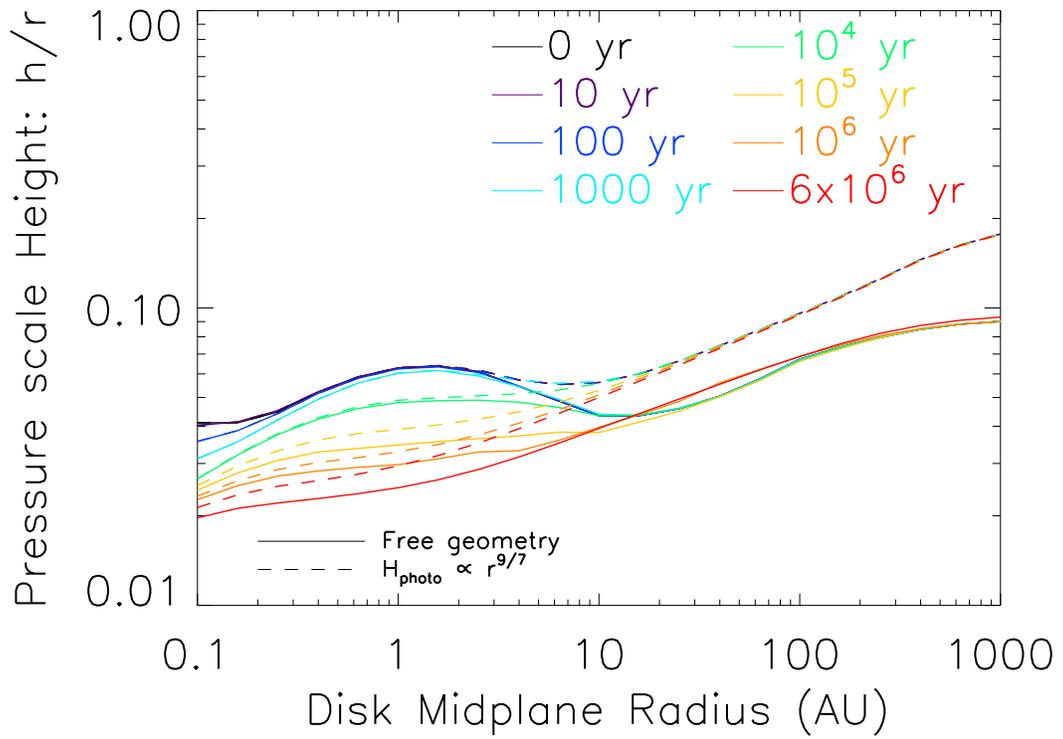}
\caption{Pressure scale height profile evolution for a Minimum Mass Solar Nebula in the case of a self-consistently calculated geometry and an imposed geometry following $r^{9/7}$.}
\label{mmsnhcg}
\end{figure}

Assuming the disk to be vertically isothermal, we can derive the pressure scale height from the mid-plane temperature. The evolution of this characteristic height profile is presented in Figure \ref{mmsnhcg}.

\begin{figure}[htbp!]
\begin{center}$
\begin{array}{cc}
\includegraphics[width=8cm]{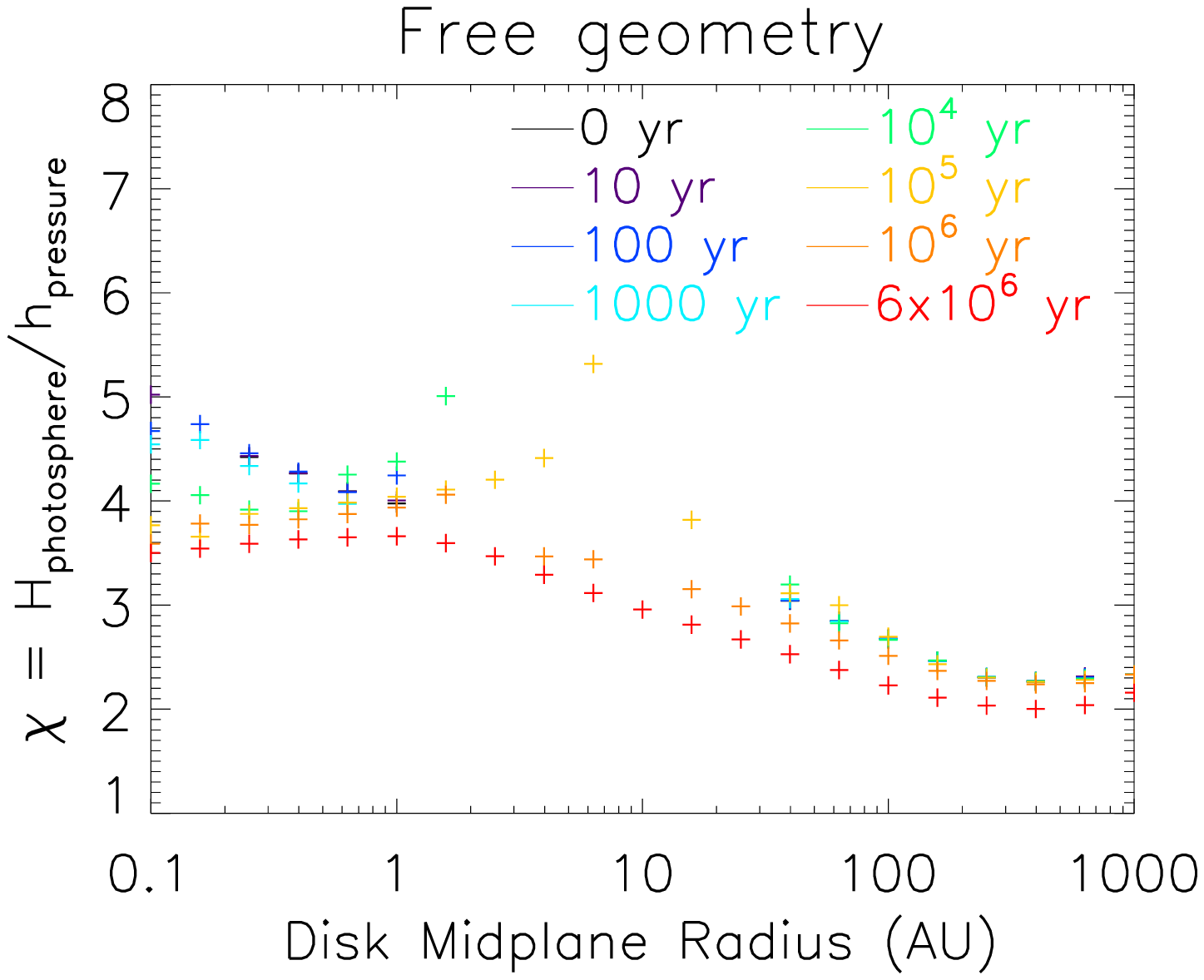}&
\includegraphics[width=8cm]{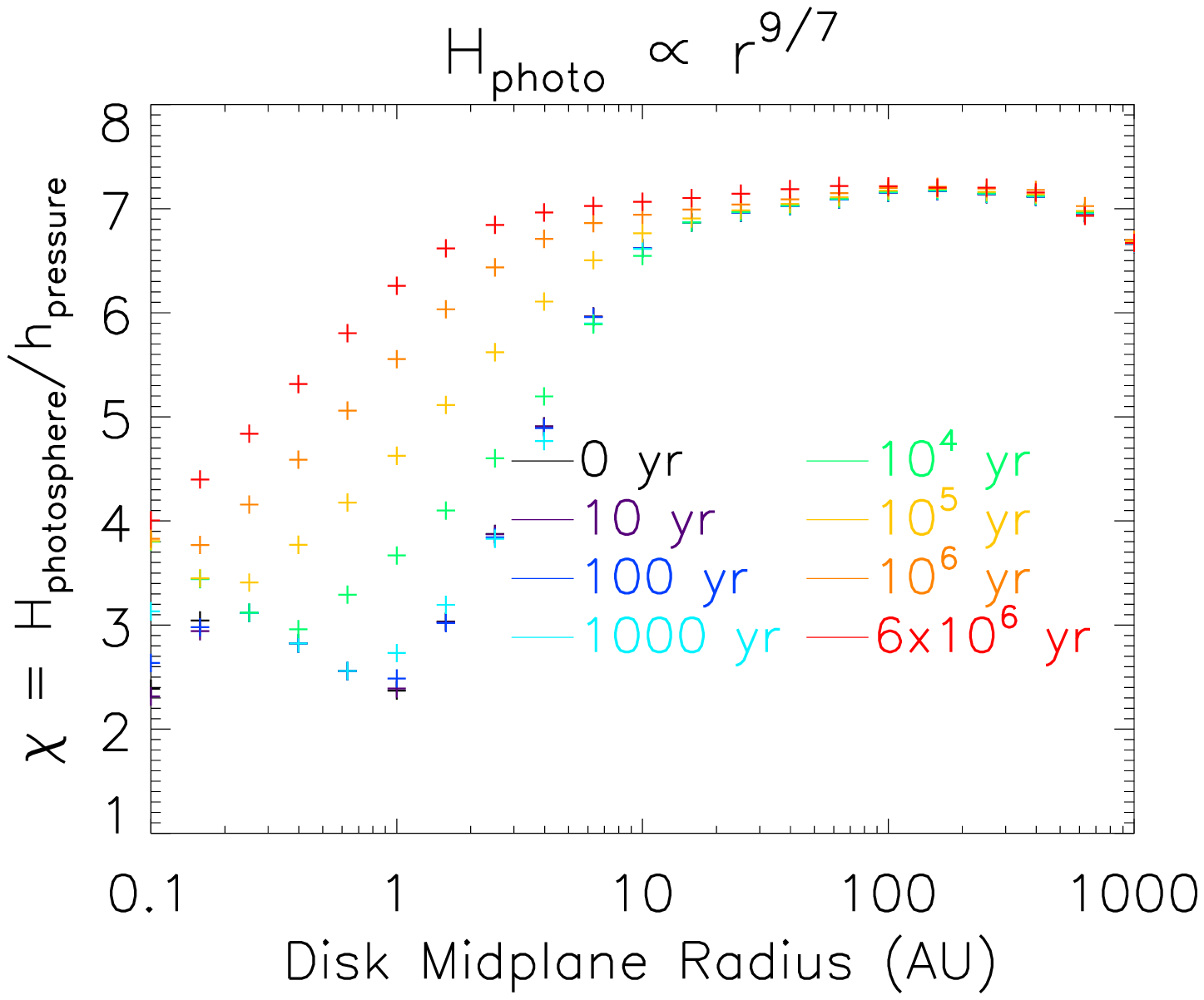}\\
\end{array}$
\caption{Pressure scale height to photosphere height ratio profile evolution for a Minimum Mass Solar Nebula in the case of a self-consistently calculated geometry (left panel) and an imposed geometry following $r^{9/7}$ (right panel). Missing points on the left panel are due to the regions of the disk being not directly in the stellar line of sight at a given location and evolution time.}
\label{mmsnchi}
\end{center}
\end{figure}

\begin{figure}[htbp!]
\center
\includegraphics[width=16cm]{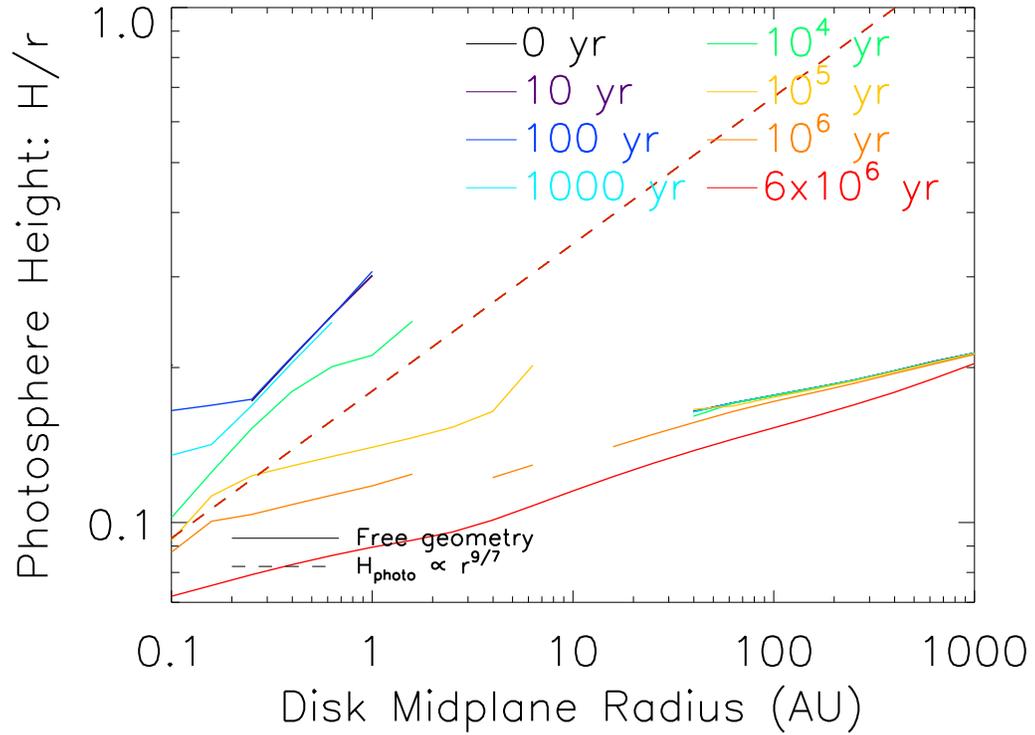}
\caption{Photosphere height profile evolution for a Minimum Mass Solar Nebula for a free geometry (solid lines). Missing points are due to the regions of the disk being not directly in the stellar line of sight at a given location and evolution time. The forced profile from \cite{chiang97} is shown in dashed line.}
\label{mmsnhphoto}
\end{figure}

In the initial state, the ratio $h_{pr}/r$ appears to be quite uniform in the inner regions and to increase slightly in the outer parts of the disk. In time, this ratio becomes a strictly increasing function of the radial distance, which asymptotic behavior can be approximated above 2 AU by a power-law $\frac{h_{pr}}{r} \propto r^{0.24}$. The derived power-law index is close to the $2/7$ value expected by \cite{chiang97} for a passive disk. In addition, the pressure scale height does not change much with time in the outer regions of the disk, which can be explained by the temperature variations being very small in the coldest and most exterior regions where the viscous heating is largely dominated by the irradiation heating (though already very small at this distance away from the star). In the case of a free geometry, the height ratio $\chi$ is only defined at the radii where the geometry will be consistent with an irradiation of the disk surface photosphere by the stellar light, i.e. when the grazing angle is actually defined. In such places, we can define a photosphere height where the optical depth reaches $1$. Evolutions of $\chi$ and the photosphere height $H_{ph}$ are presented respectively in Figures \ref{mmsnchi} and \ref{mmsnhphoto}.

Though the free-geometry $\chi$ is found between 1 and 6 as suggested in \cite{dullemond01}, it appears however that this ratio is neither radially uniform nor temporally constant, unlike \cite{chiang97} hypothesized. In addition, $\chi$ is decreasing outward and its behavior can be asymptotically approached by a power-law $\chi \propto r^{-0.12}$. Due to this decreasing, we logically observe that the photosphere height profile $H_{ph}$ is shallower than the pressure scale height profile (Figure \ref{mmsnhphoto}). We actually find $H_{ph} \propto r^{1.1}$. The ratio $H_{ph}/r$ is then strictly increasing with the radial distance, giving to the disk a flared aspect.

The internal regions are not entirely irradiated until after 100,000 years of evolution. This shadowing phenomenon appears in a similar timescale than the one required for the disk to become fully accretional and forget the initial condition on the surface mass density profile. The reached steady state presents a geometry consistent with the fact that the disk is irradiated by the star everywhere: every point of the photosphere of a flared disk is in a direct line-of-sight of the star.

\section{Discussion} \label{disc}

We have seen that, after a few million years, our disk reaches somewhat a steady state, characterized by a uniform mass accretion rate and power-law profiles of temperature and surface mass density.

\subsection{A common asymptotic state}
Based on the general trends derived from the observations, we may wonder how the initial disk is affecting the steady state if any. It appears that for a given central star with a given disk mass, different initial surface-mass distribution will lead to similar asymptotic steady states. However, the evolution timescales may vary: a disk which density is steeper will evolve slower and will unlikely reach a steady state before its evaporation, therefore keeping the memory of the initial surface mass density profile in the outer regions. Thus, since no disk has been observed with a very steep power-law, we can state that it is very unlikely that protoplanetary disks are steeper than $\Sigma \propto r^{-2}$ when they form.

Similar effect will be observed when decreasing $\alpha_{\mathrm{visc}}$. Thus, for a transitional disk, the density distribution could provide information to retrace the disk history using the position of the discontinuity in the power-law fits. Indeed, the initial conditions being relaxed inner to this radial distance, this location will constrain the evolution time elapsed, while the power-law index of the outer profile would provide a lower estimate for the initial distribution power-law index. However, an assumption on the $\alpha_{\mathrm{visc}}$ parameter will remain necessary in order to estimate the disk age as it influences critically the relaxation speed of the disk. 

Such a disk reaches a steady state for which the asymptotic trends confirm the analytical developments giving its geometry and thermodynamics (see Section \ref{ana} and \cite{kenyon87}). Indeed, with a surface mass density radial trend in $\Sigma(r) \propto r^{-1.03}$, Equation \ref{steady2} then leads to a temperature variation in $T_{m}(r) \propto r^{-0.47}$, which is close enough to the power law index of $-0.5$ observed in our simulations. The photosphere height should therefore follow a law in $H_{ph}(r) \propto r^{1.12}$, where we measure an index of $1.1$ and where \cite{kenyon87} was expecting to find $9/8$, therefore validating the hypothesis of a flared geometry. Then, from the observed trend of the pressure scale height $h_{pr}(r) \propto r^{1.24}$, we can estimate the asymptotic behavior of the ratio $\chi \propto r^{-0.12}$. In addition, we notice that the pressure scale height profile is very close to the profile expected by \cite{chiang97}: $h_{pr}(r) \propto r^{9/7}$. However, their paper considered $\chi$ to be constant and uniform, what would lead to a power-law index of $9/7$ also for the photosphere height. The fact that we did not make any assumption about setting $\chi$ explains the difference between our photosphere height power-law and the one derived by \cite{chiang97}. The viscosity $\nu$ should then have a power-law index around $1.03$ and finally, the final grazing angle would vary as $\alpha(r) \propto r^{0.12}$ verifying that the disk is completely irradiated in the steady state.

Numerical simulations from \cite{watanabe08} have shown that thermal instabilities may generate peaks in the photosphere height profile that may project shadows on the regions located directly outside these peaks. These non-irradiated zones present lower temperatures. However, these results were obtained by modeling thermal transfers in the disk while the surface mass density is kept constant, whereas in our simulations, we assume an instantaneous hydrostatic equilibrium but a viscously evolving surface mass density. Therefore, these thermal instabilities are logically not visible in our simulations. As \cite{watanabe08} detailed, variations of $\chi$ may generate instabilities, and we have shown in Figure \ref{mmsnchi} that it cannot be considered either uniform nor constant. Therefore, one would expect thermal waves to be able to appear in such disks. The apparition of quasi-periodic thermal waves seems to require mass accretion rates lower than $10^{-7} ~\mathrm{M_{\odot}\cdot yr^{-1}}$, which is only possible after 100,000 years of evolution, and limited to the region of the disk inner to 20 AU, where the thermal timescale is much shorter than the dynamical and viscous timescales. Attempting to model these instabilities with our code would require taking into account time-dependent thermal transfer, constraining the evolution time step to values lower than a fraction of the thermal wave period and increasing the radial resolution. This would obviously result in an explosion of the computation time. However, those waves mainly affect the local and instantaneous temperature, generating discrepancies of a few Kelvins. Though this temperature change does not affect much the evolution of the disk, the resulting temperature gradients could affect more deeply Lindblad and corotation torques and therefore planet migration. However, in our case, we relax the assumption of a constant surface mass density and we consider that we reach the hydrostatic equilibrium at each time step, thus averaging the effects of thermal waves over time. Though this is not investigated any further in the present paper, this will certainly deserve to be developed in a future article.

The influence of the star and the disk mass will be the object of a more thorough investigation in a future paper.

\subsection{Invariance with the initial conditions}
\label{ini}

\begin{figure}[htbp!]
\begin{center}$
\begin{array}{cc}
\includegraphics[width=8cm]{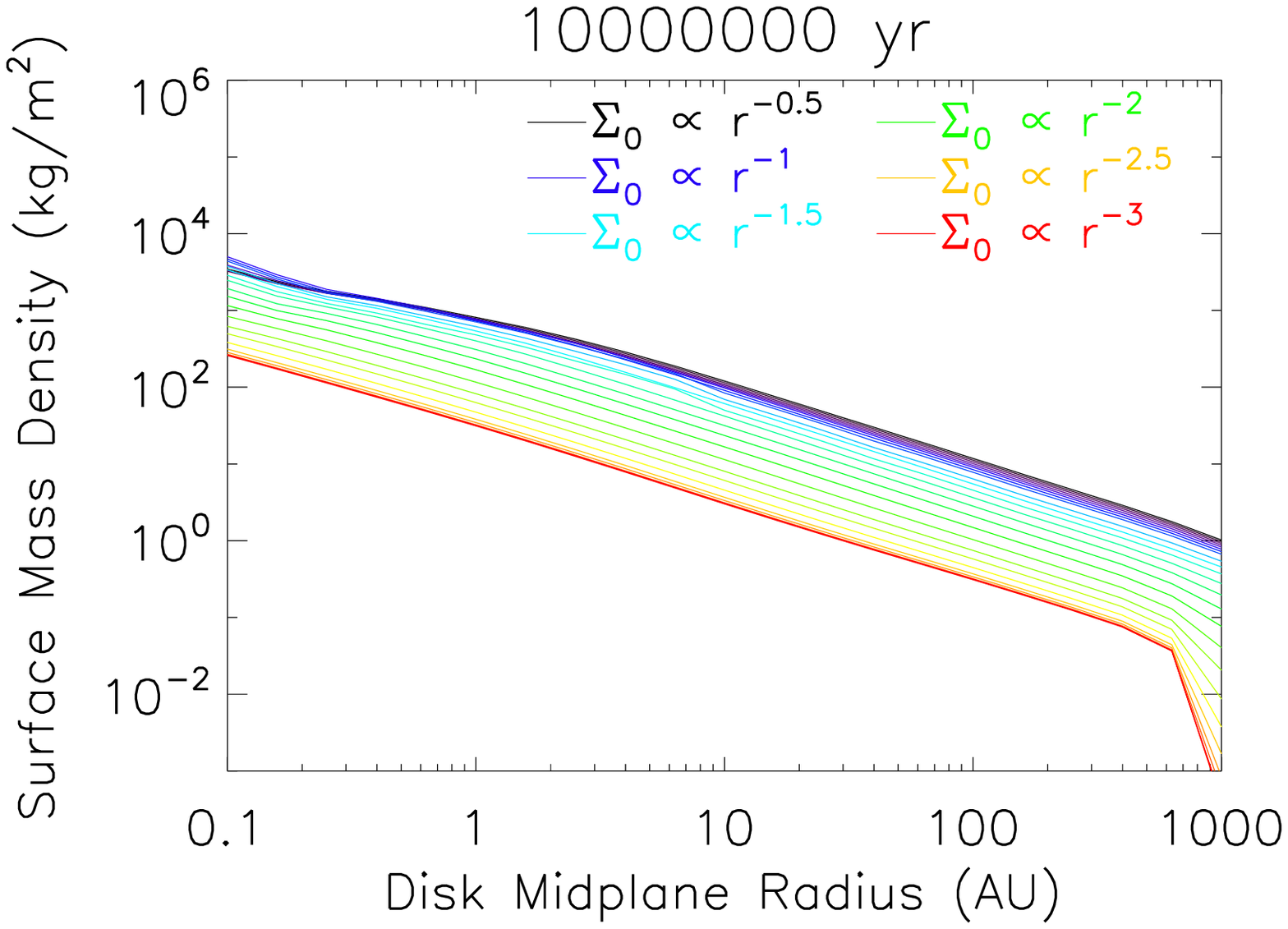}&
\includegraphics[width=8cm]{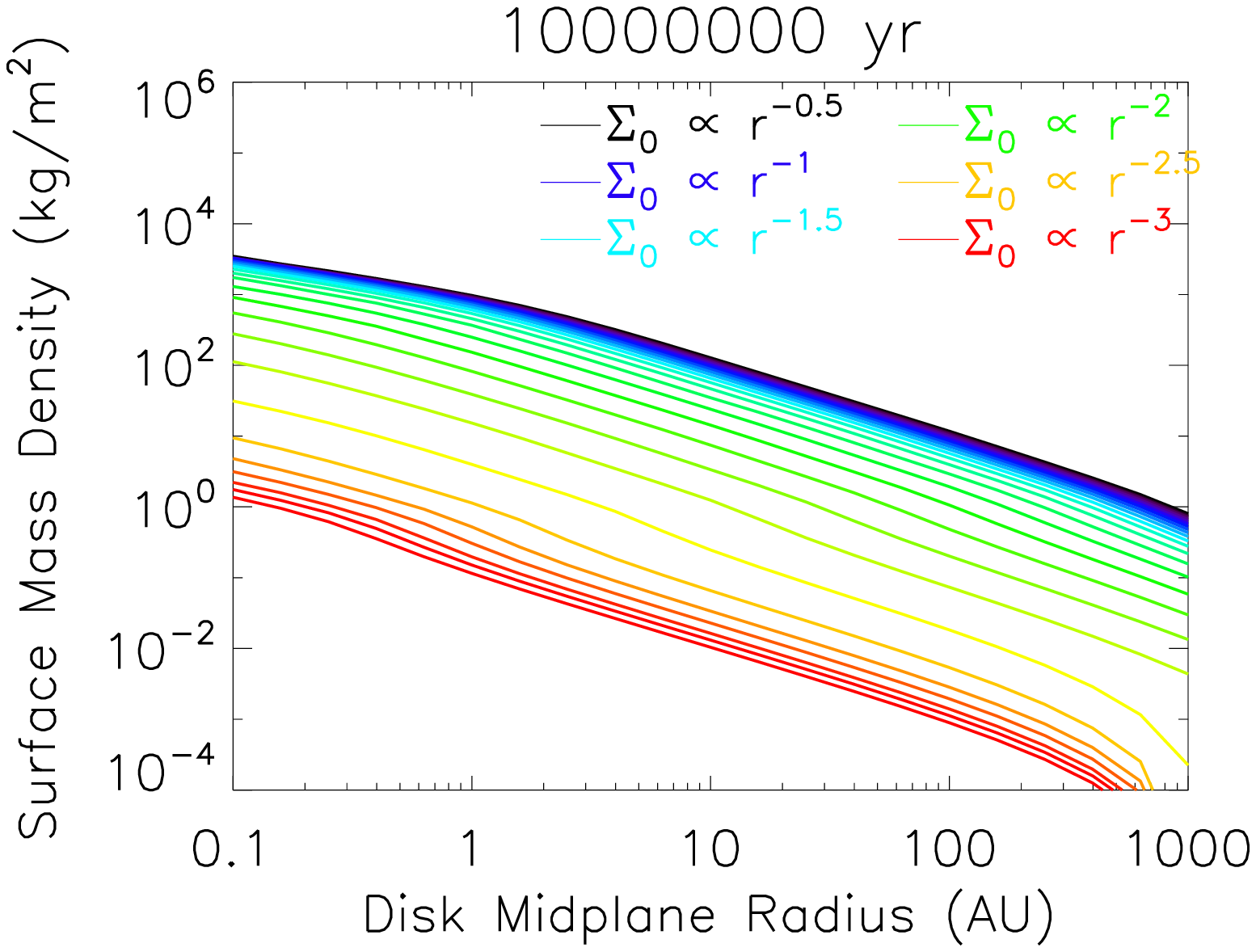}\\
\end{array}$
\caption{Surface mass density profiles for various initial conditions after 10 Million years for a self-consistently derived geometry (left panel) and for an imposed geometry following $r^{9/7}$ (right panel).}
\label{qrsigma}
\end{center}
\end{figure}

\begin{figure}[htbp!]
\begin{center}$
\begin{array}{cc}
\includegraphics[width=8cm]{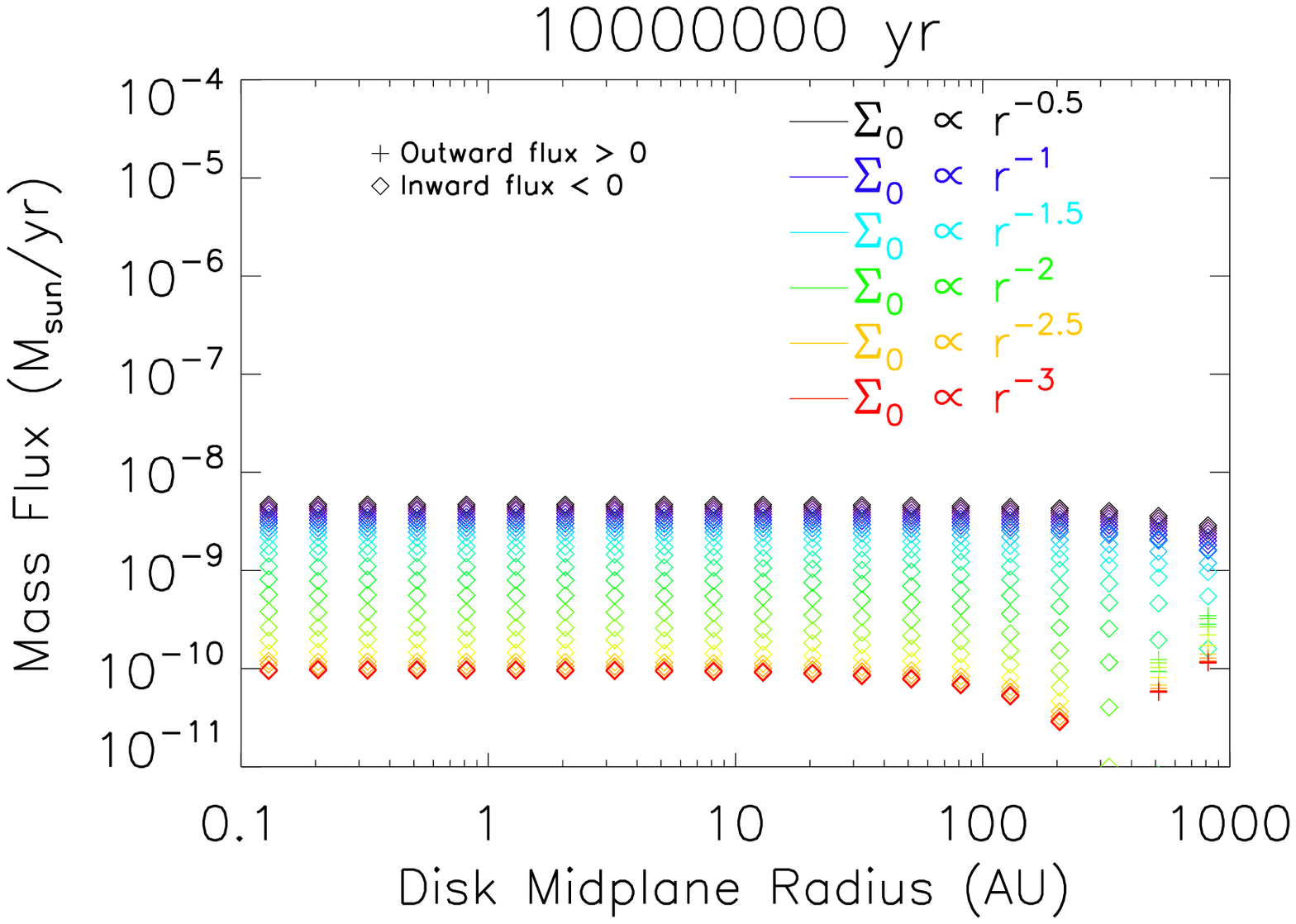}&
\includegraphics[width=8cm]{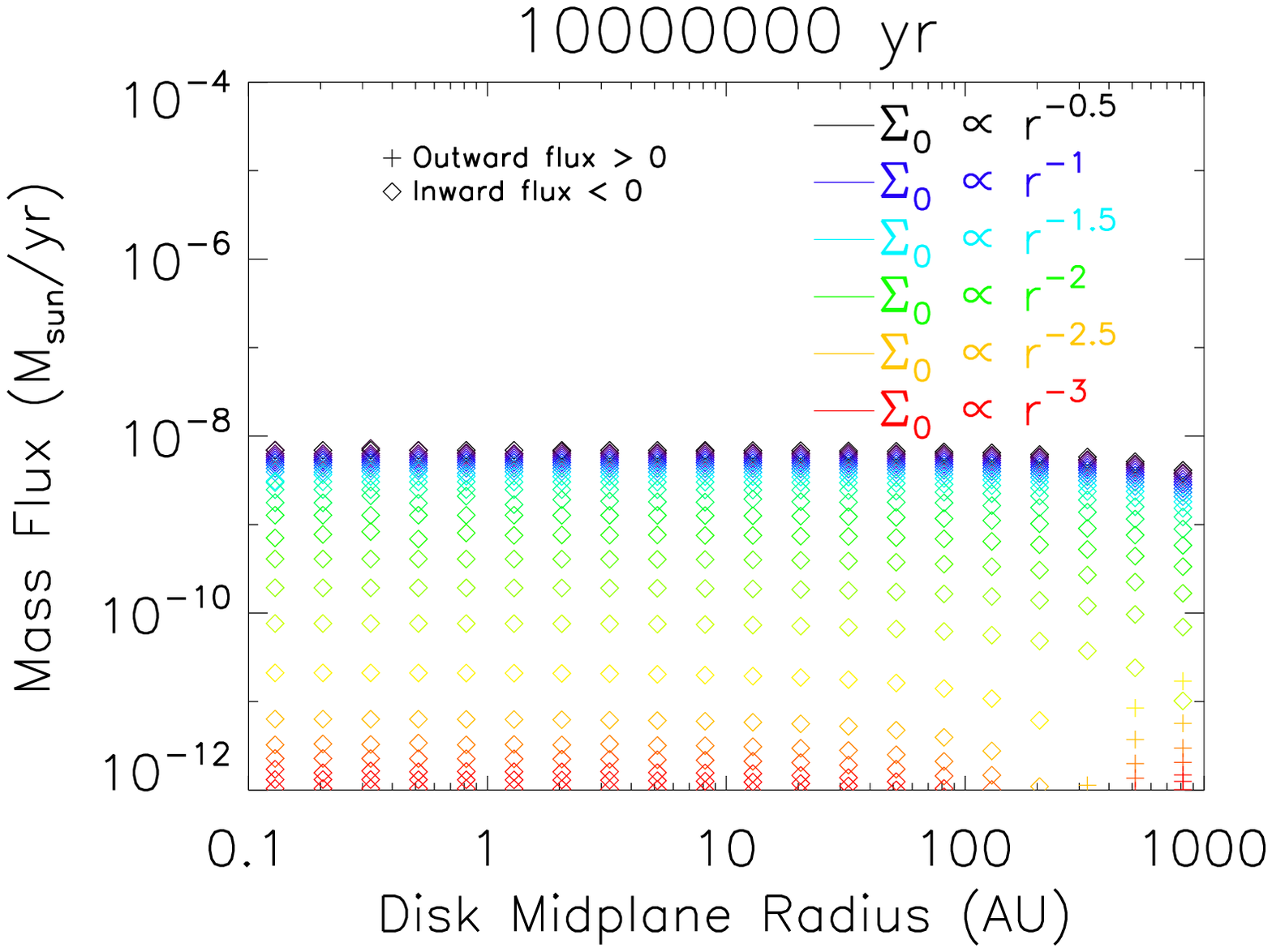}\\
\end{array}$
\caption{Mass flux profiles for various initial conditions after 10 Million years for a self-consistently derived geometry (left panel) and for an imposed geometry following $r^{9/7}$ (right panel). The steepest initial disks are not yet entirely accretional everywhere.}
\label{qrflux}
\end{center}
\end{figure}

\begin{figure}[htbp!]
\begin{center}$
\begin{array}{cc}
\includegraphics[width=8cm]{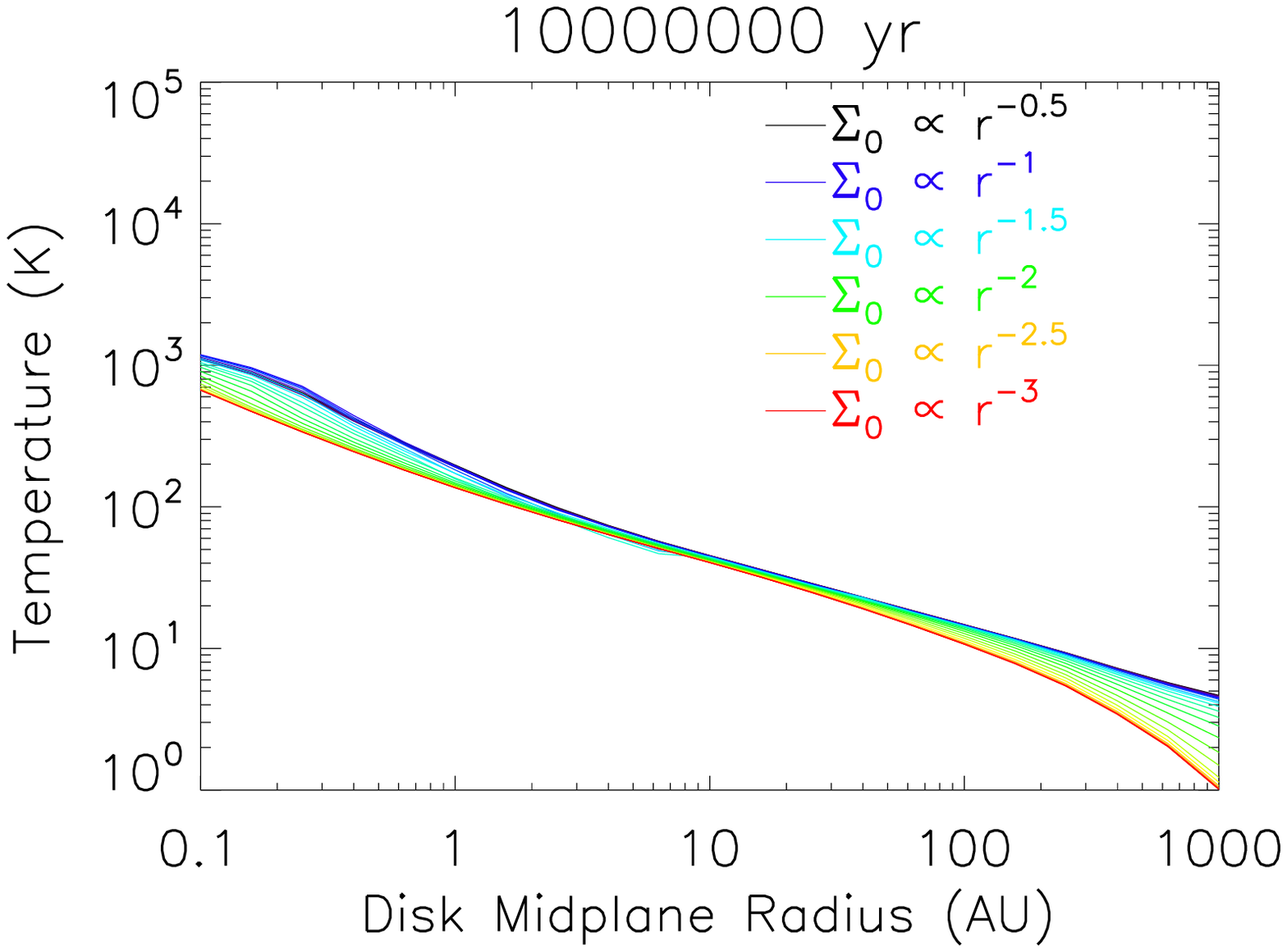}&
\includegraphics[width=8cm]{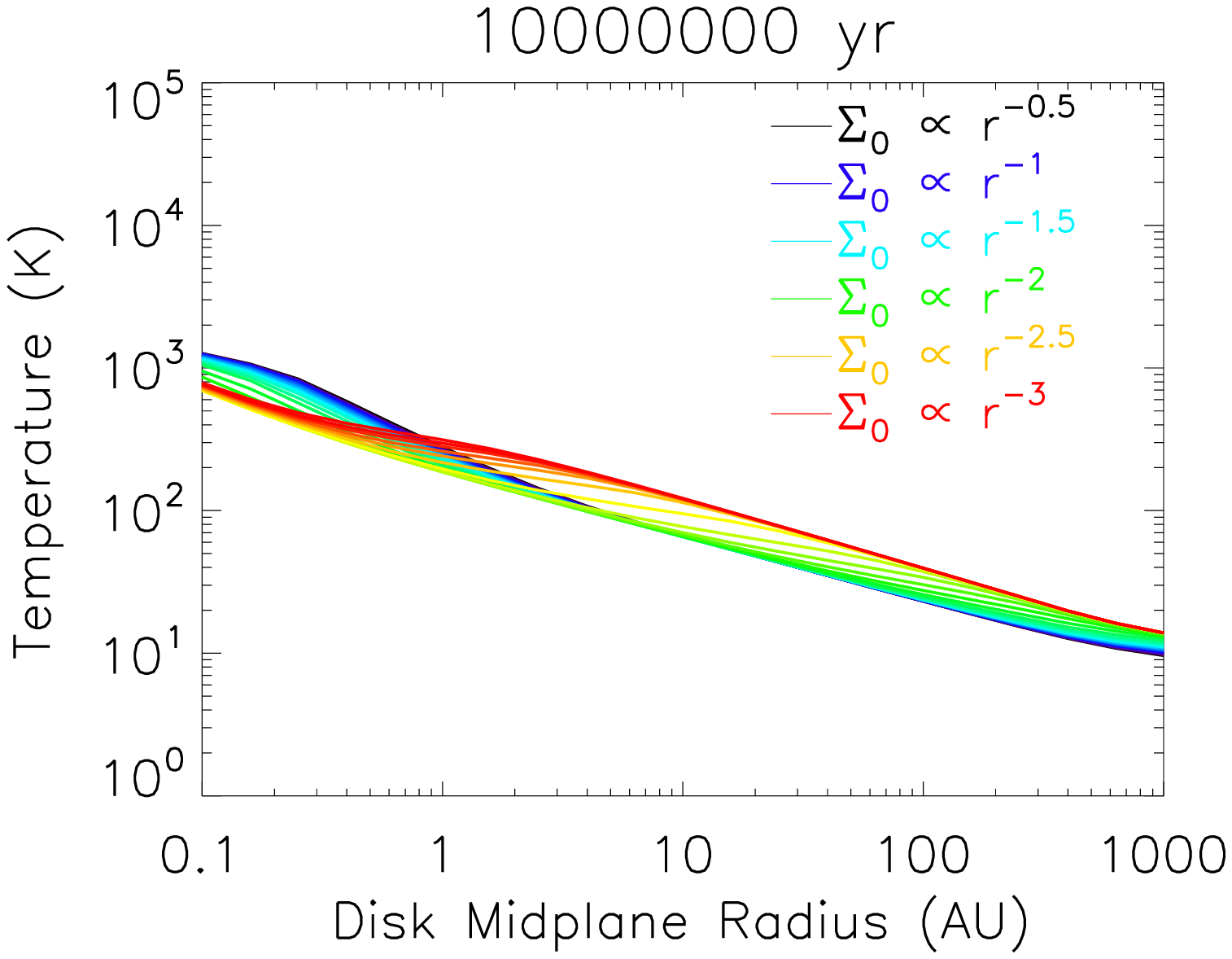}\\
\end{array}$
\caption{Temperature profiles for various initial conditions after 10 Million years for a self-consistently derived geometry (left panel) and for an imposed geometry following $r^{9/7}$ (right panel).}
\label{qrtemp}
\end{center}
\end{figure}

\begin{figure}[htbp!]
\begin{center}$
\begin{array}{cc}
\includegraphics[width=8cm]{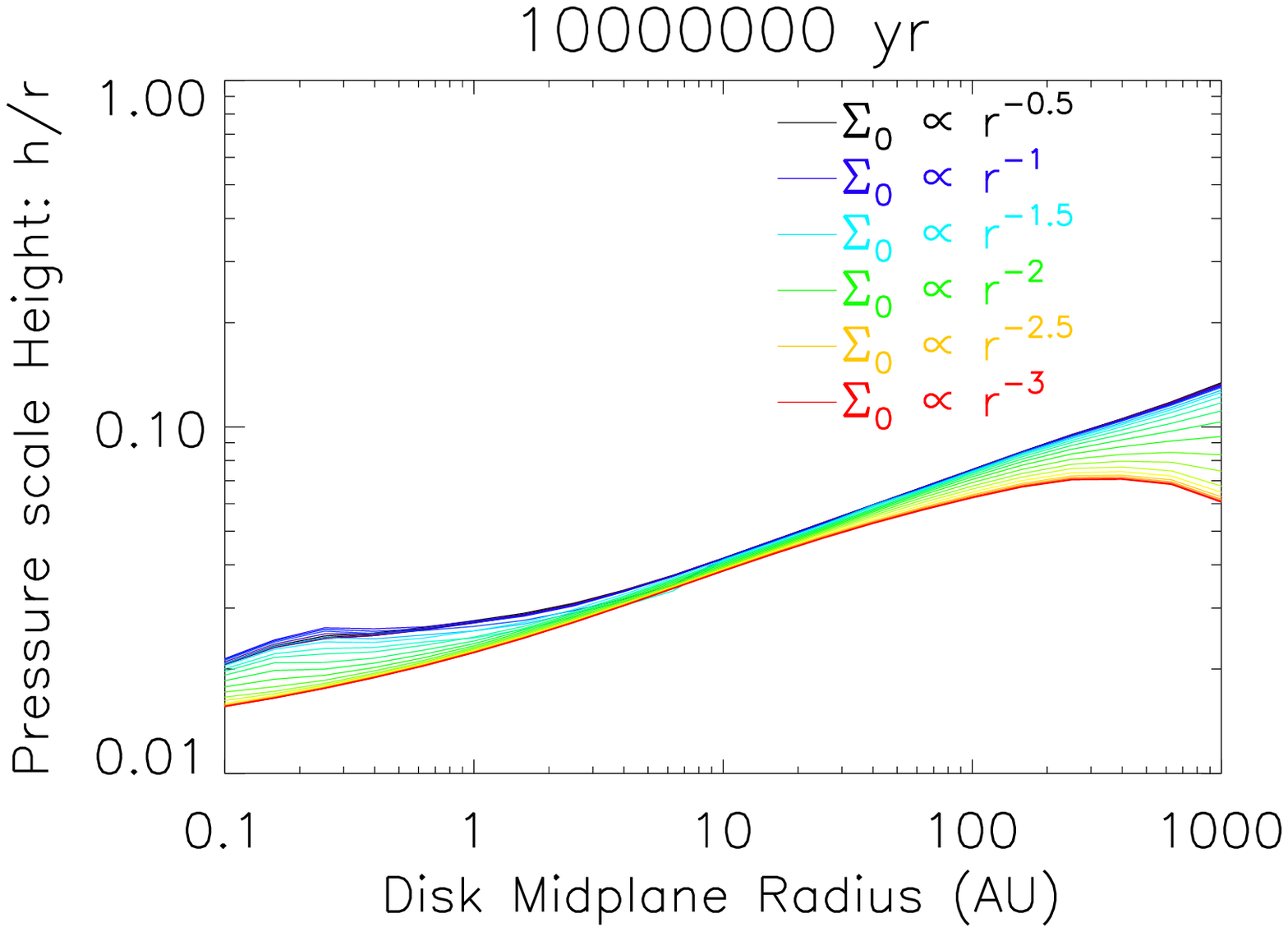}&
\includegraphics[width=8cm]{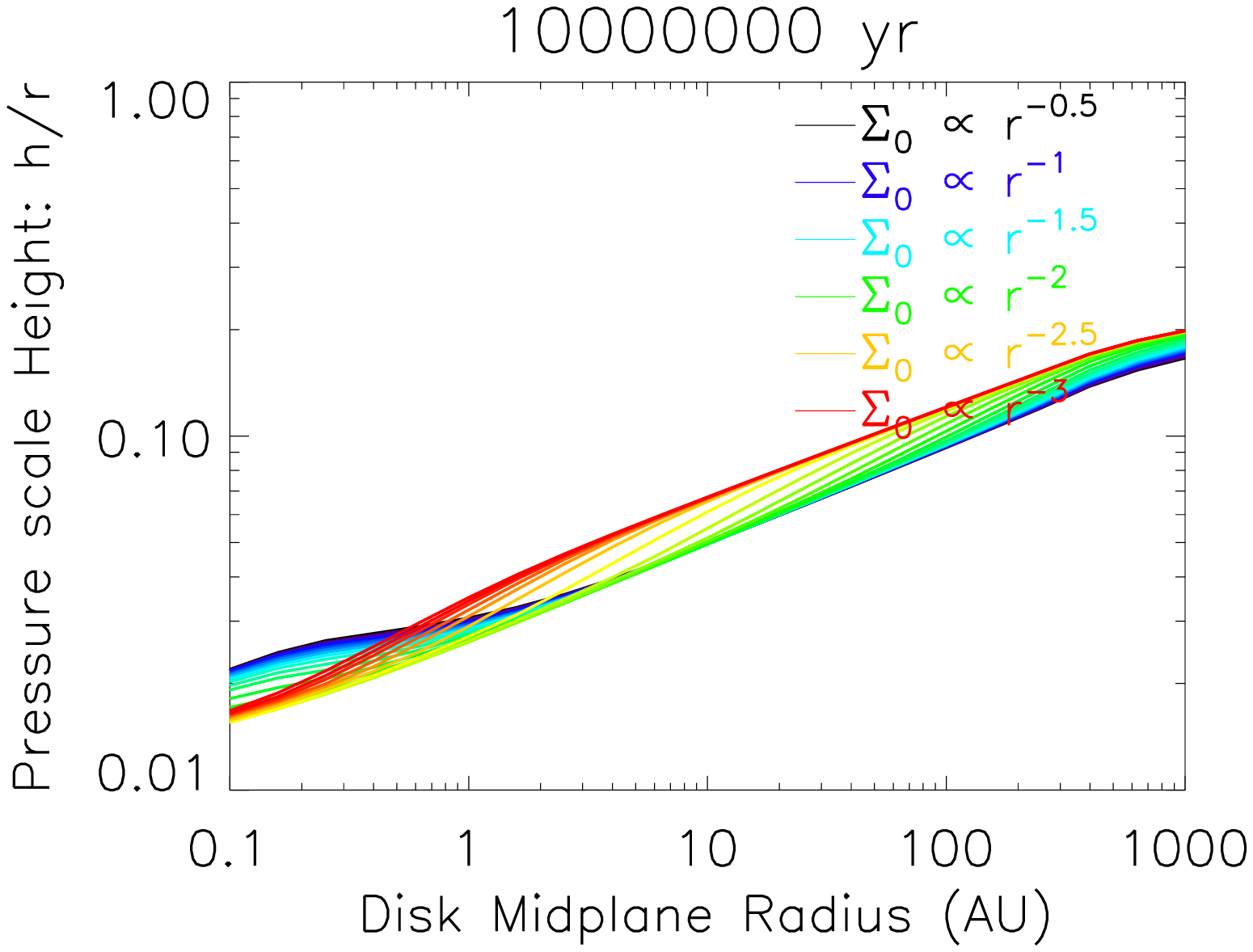}\\
\end{array}$
\caption{Pressure scale height profiles for various initial conditions after 10 Million years for a self-consistently derived geometry (left panel) and for an imposed geometry following $r^{9/7}$ (right panel).}
\label{qrhcg}
\end{center}
\end{figure}

\begin{figure}[htbp!]
\center
\includegraphics[width=16cm]{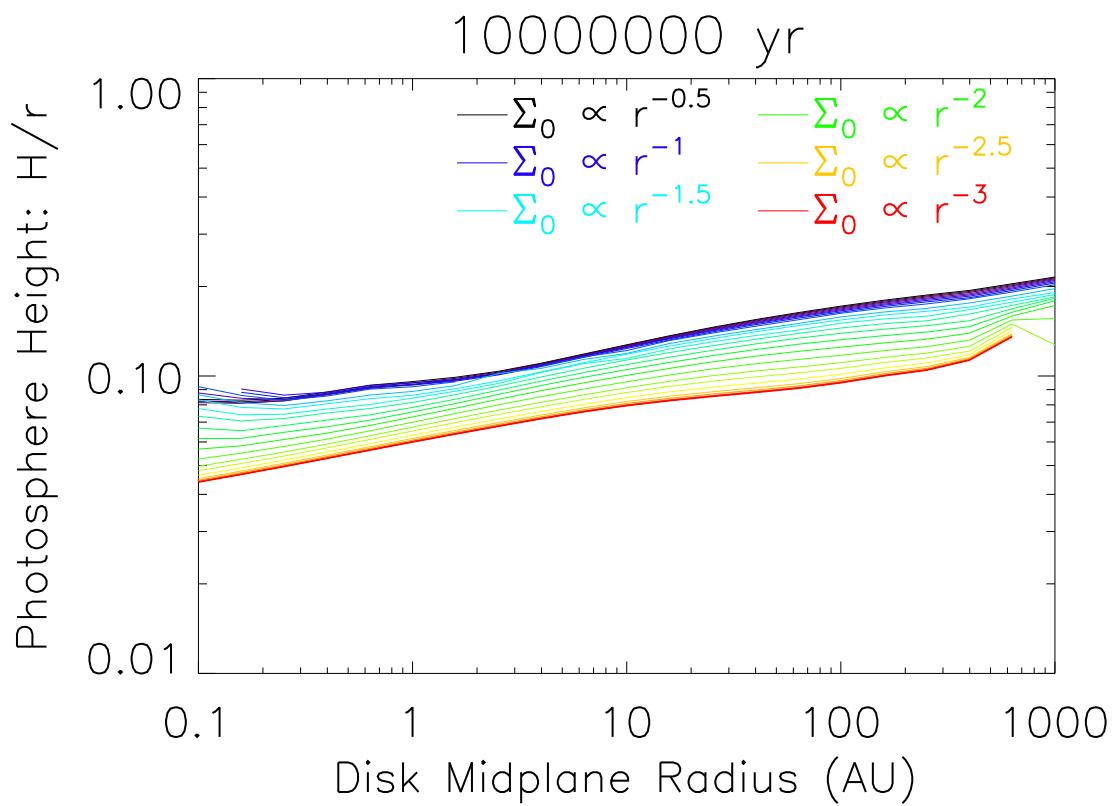}
\caption{Photosphere height profiles for various initial conditions after 10 Million years for a self-consistently derived geometry. Missing points are due to the regions of the disk being not directly in the stellar line of sight at a given location and evolution time. Irregularities at the outer edge for the steepest initial profiles are due to the optical depth being lower than 1, therefore making the photosphere height definition impossible.}
\label{qrhphoto}
\end{figure}

From Figure \ref{mmsnsigma}, we can estimate that the initial power-law index in surface mass density in the inner regions is forgotten in a few thousand years only. The final trend in $\Sigma \approx \Sigma_{1} \cdot r^{-1}$ is reached fairly early and the later evolution only results in a damping of the power-law amplitude $\Sigma_{1}$. At each moment, we can define a radial limit for which the inner density has reached a steady state power-law index, while the outer density remains quite unaffected. It appears to take up to a few million years before the outer regions reach the steady state. It is very important to understand the impact of the choice of the initial condition over these characteristic timescales and the steady state in order to estimate the validity of the conclusions of the evolution. For the same central star, we present the evolution of different initial surface mass density profiles. The disks shown in Figures \ref{qrsigma}-\ref{qrhphoto} follow an initial surface mass density profile in $\Sigma(r) \propto r^{q_{m}}$ with $-3 \le q_{m} \le -0.5$. These disks all have the same total mass as the MMSN: $\int_{0.01~AU}^{1000~AU} 2 \pi r \cdot 17,000 \cdot \left(\frac{r}{AU}\right)^{-1.5} dr = 7.5 \times 10^{-2} \mathrm{M_{\odot}}$; and their angular momentum goes from $8.1 \times 10^{13} \mathrm{kg \cdot m^{2} \cdot s^{-1}}$ for $q_{m} = -3$ to $8.0 \times 10^{15} \mathrm{kg \cdot m^{2} \cdot s^{-1}}$ for $q_{m} = -0.5$.

Figure \ref{qrsigma} shows that steeper or shallower initial disks will tend to the same surface mass density profile after a few million years. We retrieve, in every case, the final asymptotic trend in $\sim ~r^{-1}$ after 10 million years of evolution, whereas the final amplitudes tend to decrease when the disk gets steeper. Figures \ref{qrsigma}-\ref{qrhcg} provide comparison with the imposed geometry evolution. However, due to numerical limitations, those long-time simulations were limited to $q_{m} \ge -2.5$.

We also notice that the accretion rates tend to uniform profiles with lower amplitudes for steeper initial disks (Figure \ref{qrflux}). This is consistent with final surface mass densities being lower for steeper initial disks (i.e. lower angular momentum): indeed, the steeper the disk, the faster it will empty at the beginning of its evolution and the less massive the remaining disk, leading to lower accretion rates when the steady state is reached. Moreover, we notice that the steady state is not reached in 10 million years for the most compact disks (steepest power-law profiles with $q_{m} \le -2.2$): the outermost regions of the disk for $q_{m} = -3$ is still spreading outward at this date.

The dispersion between the temperature profiles for various initial $q_{m}$ is very narrow after 10 million years (Figure \ref{qrtemp}). While the inner regions show a dispersion of a few tens of Kelvins, it drops to a few Kelvins above 2 AU. The temperature profile between 1 AU and 100 AU may be approached with $T_{m}(r) ~ \propto ~ r^{-0.5}$.

However, we observe a dispersion of a few Kelvins in the 100 - 1000 AU region, where the temperature is so low that it would require more thorough model to constrain it above the CMB temperature and remove the effects of the outflow boundary conditions. Despite this variation is small in amplitude, this affects the fit of the temperature profile by a power-law. This effect is also quite sensitive in the pressure scale height (Figure \ref{qrhcg}) and photosphere profiles (Figure \ref{qrhphoto}). We suggest that this dispersion is explained by a slower relaxation of the initial condition in the outer parts for the steepest initial disks that are not yet completely irradiated and flared everywhere after 10 million years. This is due to the fact that these disks have more material in the inner regions and therefore require a longer time to reach their steady state. However, despite various initial conditions, we find that these disk photospheres evolve toward a common asymptotic trend between 1 AU and 100 AU, characterized by:
\begin{eqnarray}
h_{pr}(r) &\propto& r^{1.23 \pm 0.02}\\
H_{ph}(r) &\propto& r^{1.10 \pm 0.02}
\end{eqnarray}

\subsection{The importance of a realistic geometrical structure}
\label{geo}

\cite{chiang97} provided a semi-analytical expression for the photosphere height in the case of a passive disk that has reached its steady state. \cite{hueso05} and \cite{birnstiel10} used this result as a prescription to constrain their geometry while \cite{brauer08} imposed the grazing angle to have a fixed value. Though these geometries may fit in the case of passive disks, this is only working for viscous heated disks in the outermost regions, where the viscous heating is totally dominated by the irradiation heating. However, the scope of our simulations is much larger than that and our code allows studying the transitory states of the disk and especially the intermediate and inner regions where planets are thought to form. Therefore we can only compare our results to the \cite{chiang97} model in the outer zones and after a time long enough for the steady state to be reached.

Thus, looking at the outer parts of the disk after a few million years of evolution, we notice that the surface mass density profile does not seem significantly different, which can be understood since the outer regions of the disk keep for a longer time the memory of the initial conditions. The temperature profile on the other hand is clearly lower (lower contribution from the irradiation heating) and slightly steeper, consistent with lower pressure scale and photosphere heights. These heights are slightly shallower than in the forced geometry model but present asymptotic power-law index quite close to the ones derived by \cite{chiang97}. However, the geometry of the transition zone is clearly different and the height ratio $\chi$ is definitely neither constant nor uniform.

The forced geometry model used in previous works is clearly a good approximation for global long-term evolution. However, it is not adapted for the study of smaller structures such as shadowed regions, transition zones between viscous and irradiation domination or the regions where changes of states may occur (such as the sublimation zones). Modeling these phenomenons requires the geometric refinement we suggest in order to generate irregularities in surface mass density or temperature that are thought to favor the formation of planetary traps where planetary cores might accrete. This will be the purpose of a future study. Shadowing may also help explain observed structures in transition disks: \cite{siebenmorgen12} showed that temperature drops due to shadows may generate dark bands in mid-infrared observations at 10 $\mathrm{\mu m}$.

\subsection{Comparison to observations}
\label{obs}
Since the steady state appears in a time much shorter than the gas disk typical lifetime, we may assume that observed disks are more likely to be observed in their steady state phase. Therefore, we investigate how our results may reproduce the observations of disk surface-densities, shapes and accretion rates. Recent works from \cite{andrews09,andrews10} and \cite{isella09} reported observations of the Ophiuchus and Taurus regions and derived power-law fits of the disk photosphere heights and surface mass densities, as well as accretion rates for young stars.

\cite{andrews09, andrews10} reported that the quasi totality of their surface mass density fits belonged to the range [0.8,1.1]. These results compare well with the asymptotic states obtained in our simulations where we find an index of $\approx -1$ (see Figure \ref{qrsigma}).

We find that accretion rates are also concurring with measured values: \cite{hartmann98}, \cite{andrews09,andrews10} and \cite{isella09} reported $10^{-10} ~\mathrm{M_{\odot}\cdot yr^{-1}} \le \dot{M} \le 10^{-7} ~\mathrm{M_{\odot}\cdot yr^{-1}}$ in the Taurus and Ophiuchus young star regions.

\cite{andrews09} found power-law indices for the photosphere height between $1.06$ and $1.15$ for disks without inner cavities. This study was confirmed for fainter sources of the same region by \cite{andrews10} who found that most of the disks without cavities are in the same range.

Finally, \cite{lagage06} observed the disk around HD97048, a young intermediate-mass star, and found $H_{ph} \propto r^{1.26 ~\pm~ 0.05}$. Therefore this system could be one example of a disk with a constant ratio of photosphere height and pressure scale height at all times as assumed by \cite{chiang97}. \cite{andrews09} reported similar observations for the disk around SR21 that presents an inner cavity: they found $H_{ph}(r) \propto r^{1.26}$. Since these two observations do not seem to be possible to obtain from the evolution of our simple disk models, it may be possible that these disks have more complex structures, especially internally.

\subsection{Early evolution characteristics}
\begin{figure}[htbp!]
\begin{center}$
\begin{array}{cc}
\includegraphics[width=8cm]{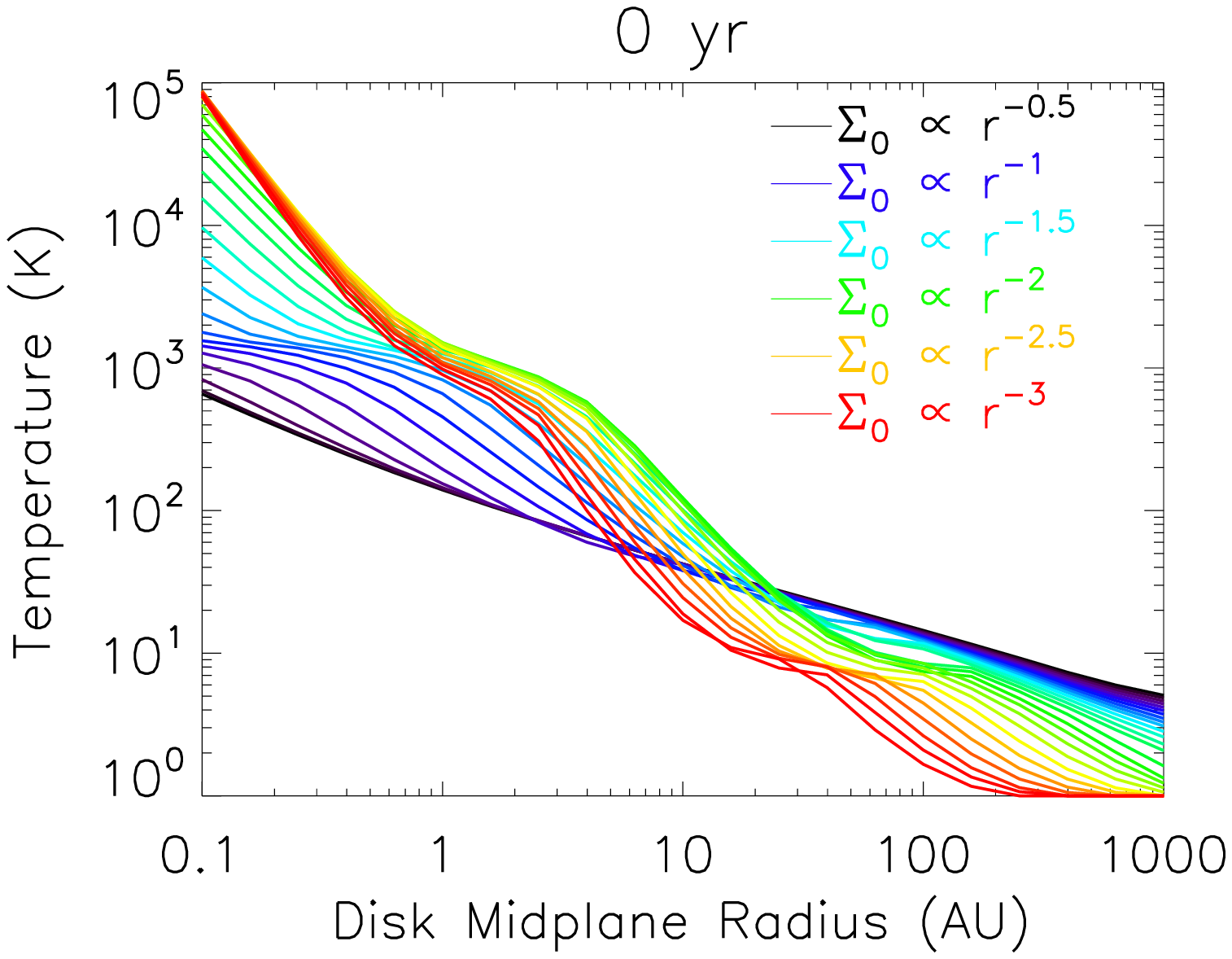}&
\includegraphics[width=8cm]{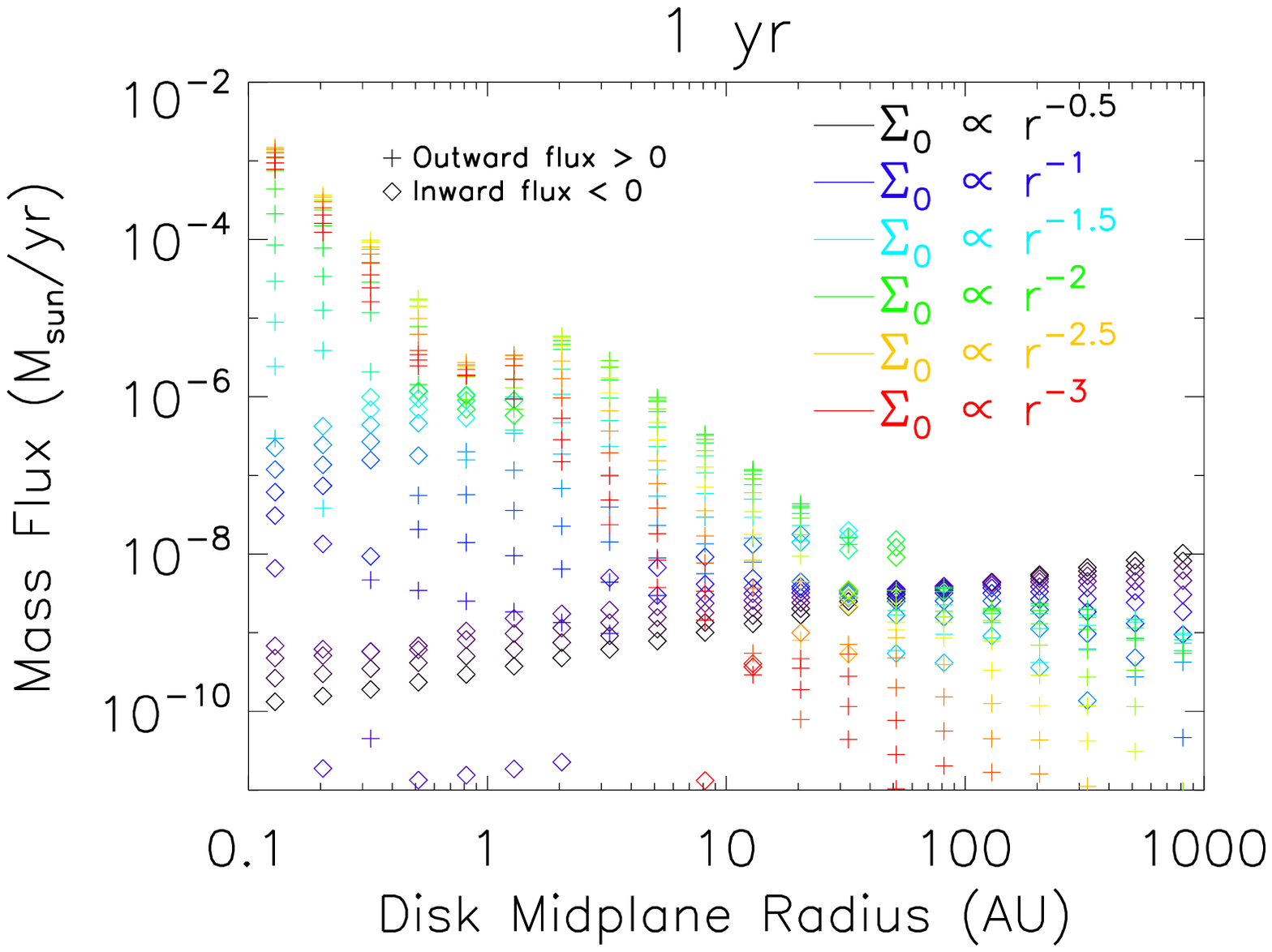}\\
\includegraphics[width=8cm]{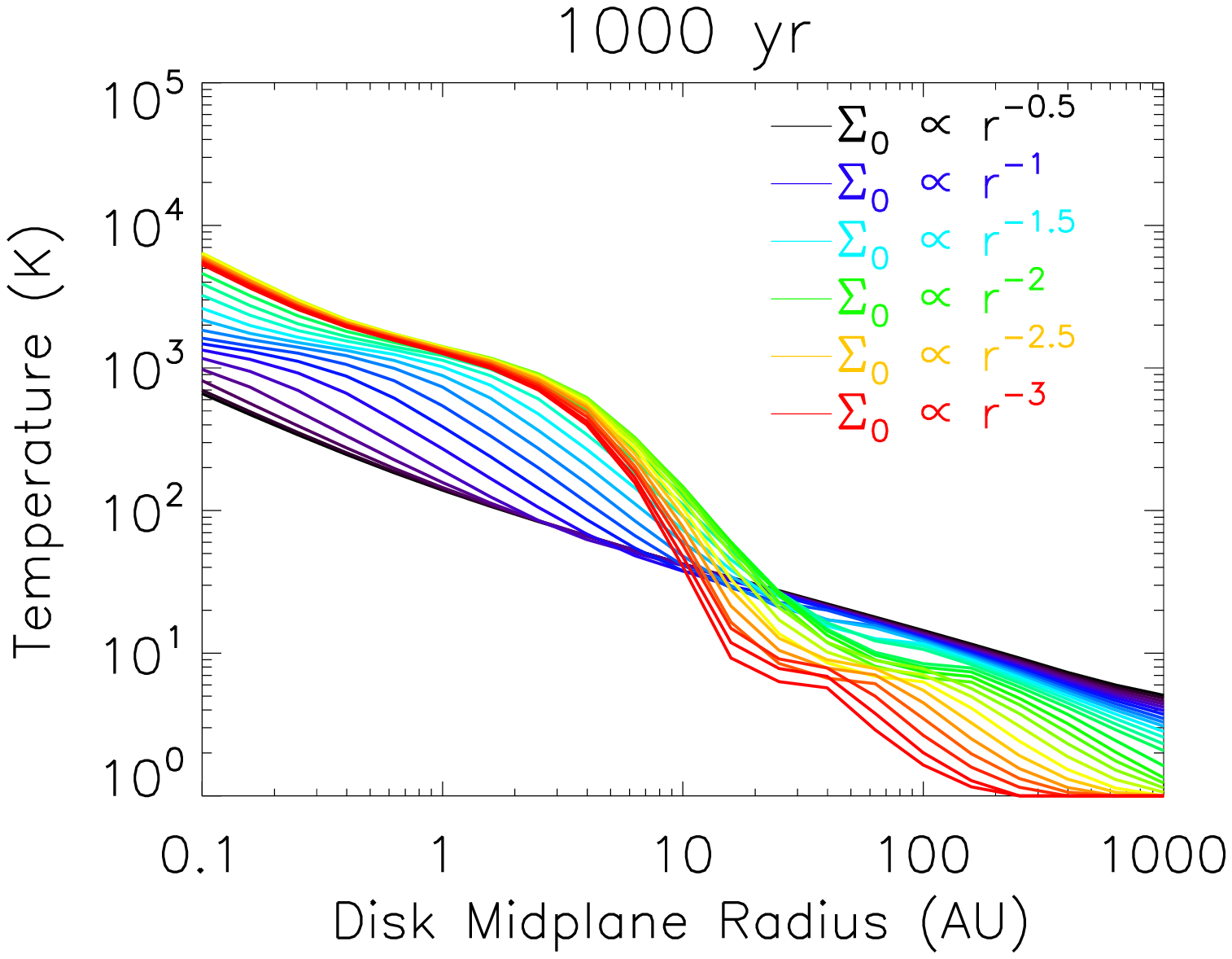}&
\includegraphics[width=8cm]{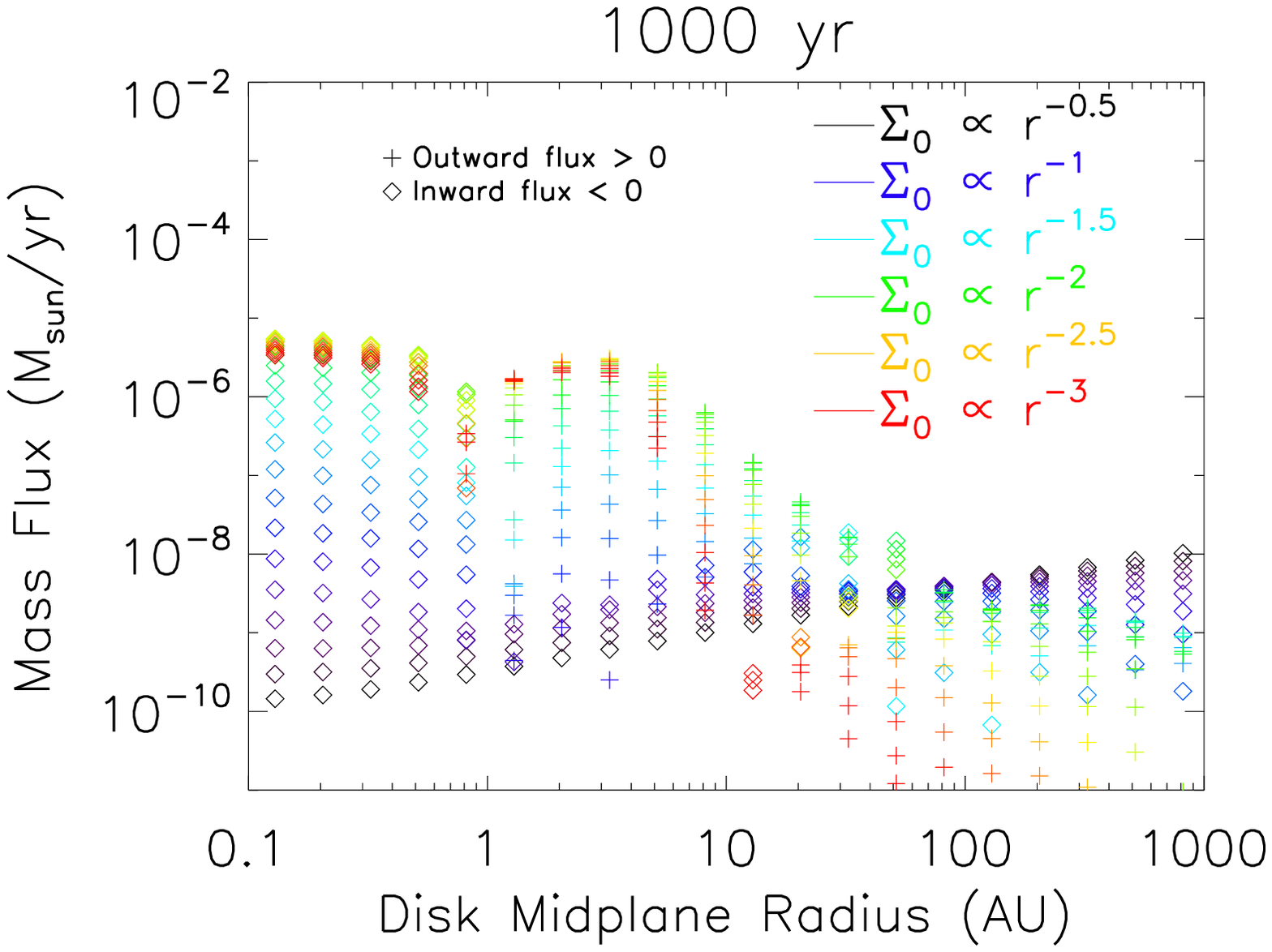}\\
\includegraphics[width=8cm]{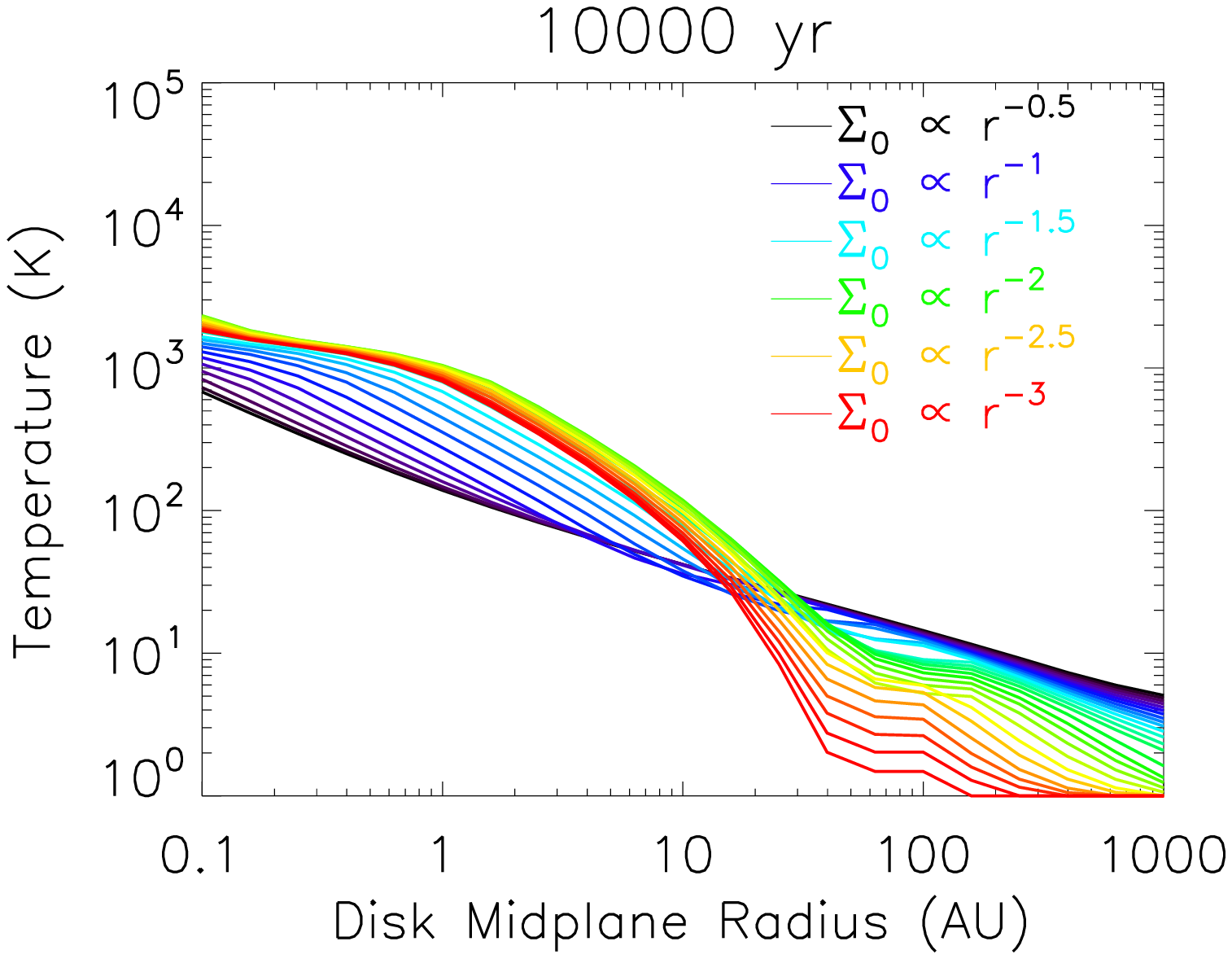}&
\includegraphics[width=8cm]{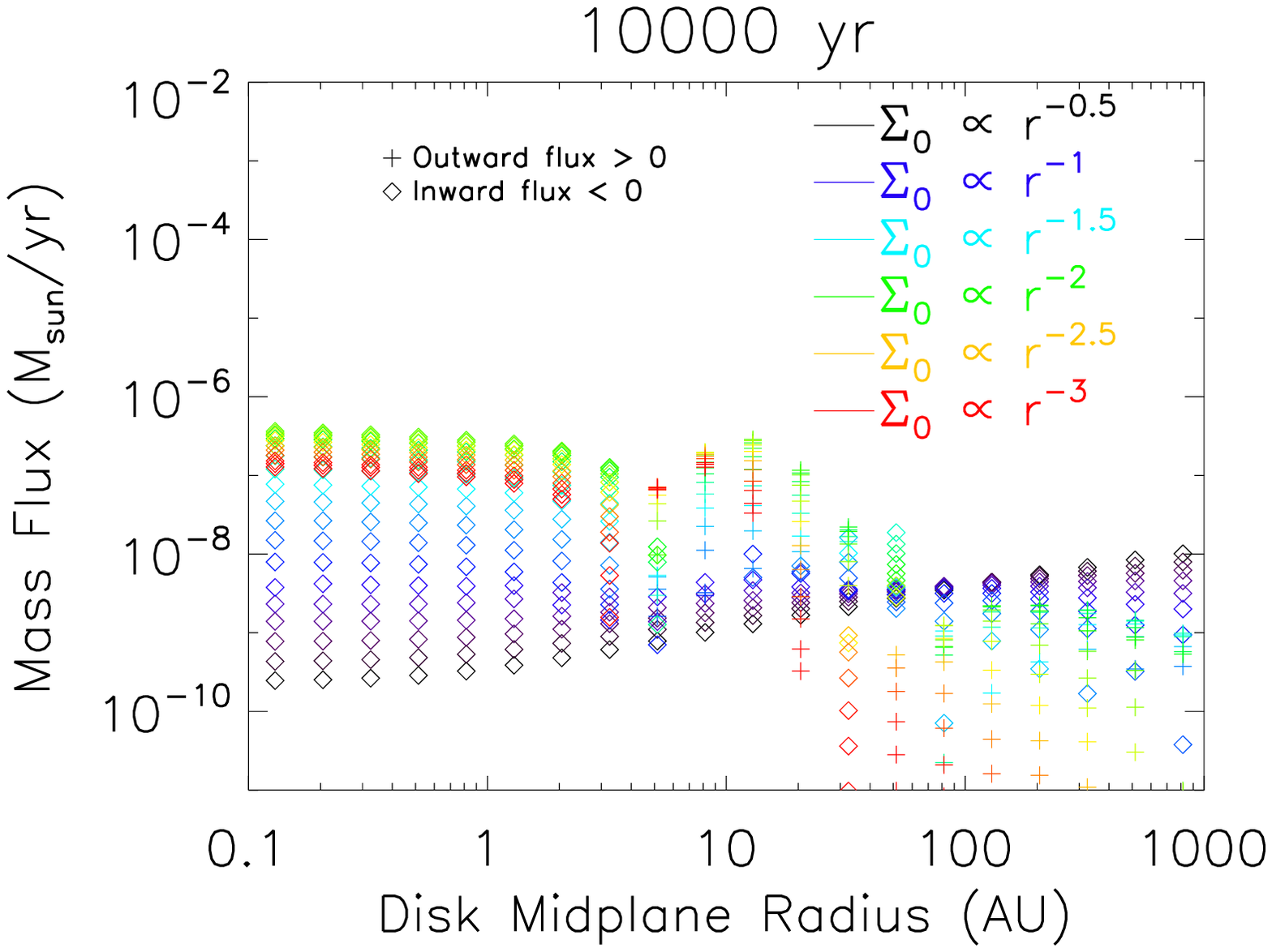}\\
\end{array}$
\caption{Snapshots of temperature and flux profiles at different evolution times for different initial surface mass density profiles.}
\label{qrcai}
\end{center}
\end{figure}

The formation of hot minerals is a critical stage in planetary formation that is expected to occur early in the disk evolution. Therefore, it is very interesting to check if the required conditions for their formation can be met in our modeled disk evolution. Starting from a primordial nebula, somewhat legitimized by \cite{vorobyov07}, it is interesting to notice that in the first 100,000 years of evolution, the temperature will remain quite hot in the inner regions. Actually, we can find a sublimation zone in the mid-plane in the first 100,000 years. In addition, the whole disk is spreading outward. As we have seen in Section \ref{sectionmmsnflux}, this flux will change direction a few tens of years later at 0.1 AU. The coexistence of these two observations shows that it is actually possible, in the first instants of evolution, to heat the dust grains enough to sublimate them, while the outward flux allows to transport them to the outer cooler regions where they can cool down, condense and form CAIs. The non-importance of the choice for the initial surface mass density is shown in Figure \ref{qrcai}: for any initial power-law surface mass density profile (with $q_{m} \ge -2$, there exists a time in the first 10,000 years of evolution that verifies these two conditions. Therefore, this result is robust to the choice of the initial surface mass density power-law index. These processes will be the objects of more thorough developments in Taillifet et al., 2013 (in preparation). This initial viscous expansion of the disk has been already observed (for example in \cite{hughes10} in which the temperature profile is imposed and constant with time, contrary to the present study).

\section{Conclusions}

Using a 1D viscous hydrodynamical code of disk viscous evolution coupled with a thermodynamical model of viscous and radiative heating, we have studied the evolution of a protoplanetary disk around a T Tauri type star. Special care was given to enforcing a consistency between the photosphere geometry and the disk thermal structure. Applying this code to a disk surrounding a young solar type star, we were able to retrieve the main characteristics of observed protoplanetary disks as reported in \cite{andrews09} and \cite{isella09}, thus validating our numerical code for further developments. We were able to characterize the steady state that appears in a few million years and retrieve its properties: a radially uniform mass flux with values matching the observed mass accretion rates. This steady state is observed despite a wide range of initial conditions and systematically tends to a surface-mass density profile varying as $r^{-1}$, no matter the initial power-law index of the density distribution. This power-law index is reached in a time that can be compared to the disk lifetime between 10 and 100 AU and in a few million years beyond 100 AU. This slower evolution in the outer regions may actually allow to trace the initial conditions as they are relaxed much later at the outer edge.

Another important result is that the disk is not always fully irradiated: shadowing can occur in the transition zone. However, after a few 100,000 years of evolution, the disk is fully irradiated, leading to a photosphere profile varying as $r^{1.1}$. This asymptotic state is a consequence of the balance between energy input due to stellar irradiation and energy loss due to viscous dissipation, resulting in a simple relation between the temperature and surface mass density power-law index (see Equation \ref{steady2}). 

This work also focused on the differences with the simple geometric models inherited from \cite{chiang97}. This geometry in $H_{photo} \propto r^{9/7}$ is actually a reliable approximation if one only focuses on the evolution of the surface mass density on a large scale. However, when interested in smaller structures, we must investigate the local geometry: we showed the importance of calculating self-consistently the geometry of the disk in agreement with its temperature. We also invalidated the approximation of a constant and uniform ratio of photosphere to pressure scale height.

While investigating the possibility to form the first solids in a protoplanetary disk, it is important to have a realistic model for the disk geometry as it will drive the temperature and surface mass density evolution. Indeed, shadowing or changes of state might create irregularities in temperature or density that are thought to be a favorable terrain to generate outward migration and therefore make planetary traps \citep{hasegawa112}.

While this study validates the detailed numerical code, we now have a precious tool to explore a huge variety of initial conditions and configurations: it is necessary to explore the stellar diversity in order to reproduce these observations individually. Our numerical model may also be improved in its physics by taking into account the feeding of the disk by the collapse of the molecular cloud for instance \citep{yang12}, improving the disk chemistry \citep{tscharnuter07}, or its dissipation by the photo-evaporation \citep{font04, alexander07, alexander09, owen10}. The internal structure may also be refined by considering the variations of the opacities with the temperature, implementing shadowing effects and variable opacities \citep{bitsch12}, or using a better Magneto-Rotational Instability model in order to add variations of the turbulent viscosity parameter and define dead-zones \citep{charnoz12,zhu101}. The stellar model may also be improved by inputing the young Sun evolution from \cite{piau11}. Understanding how the disk scales with the protostar will certainly help targeting future ALMA and JWST observations.

\acknowledgments

We thank Esther Taillifet and Eric Pantin for enlightening discussions, and the referee for detailed and constructive comments that improved the quality of the paper. This work was supported by IDEX Sorbonne Paris Cité. We acknowledge the financial support from the UnivEarthS Labex program of Sorbonne Paris Cité (ANR-10-LABX-0023 and ANR-11-IDEX-0005-02).

\appendix
\section{Asymptotic behaviors}
\label{ana}
Hereafter, we derive from basic analytical considerations of disk physics the conditions for which such a steady state is attained and compare them to our numerical results. We assume here that the temperature, surface mass density, photosphere height and pressure scale height profile asymptotic behaviors may be modeled using power-laws in the outer regions of the disk:
\begin{eqnarray}
T_{m}(r) &\propto& r^{~q_{t}}\\
\Sigma(r) &\propto& r^{~q_{m}}\\
H_{ph}(r)&\propto&r^{~q_{p}}\\
h_{pr}(r)&\propto&r^{~q_{h}}\\
\chi(r)&\propto&r^{~q_{c}}
\end{eqnarray}

We make the common assumption that in the outer regions, where the disk is the least dense, the viscous heating can be neglected compared to the stellar irradiation heating. Therefore, the mid-plane temperature in the outer part of the disk can be estimated from the irradiation alone, using Equation 1 from \cite{chiang97} considering that the flux of stellar irradiation incident upon the disk is re-radiated as a blackbody at the mid-plane temperature:
\begin{equation}
\label{tcg97}
T_{m}(r) = \left(\frac{\alpha_{gr}(r)}{2} \right)^{1/4} \left(\frac{R_{*}}{r} \right)^{1/2} T_{*}
\end{equation}

We start considering a non-flat disk ($q_{p} \ne 1$). A Taylor development of Equation \ref{alphagr} in the outermost regions allows to only keep the first term and then to write:
s\begin{equation}
\alpha_{gr} \propto (q_{p}-1) r^{q_{p}-1}
\end{equation}

Thus, Equation \ref{tcg97} provides $q_{t} = \frac{q_{p}-3}{4}$, and the pressure scale height definition gives $q_{h} = \frac{q_{p}+9}{8}$ while the expression of the ratio $\chi$ of the photosphere height over the pressure scale height leads to:
\begin{equation}
H_{ph} \propto r^{\frac{q_{p}+9}{8}+q_{c}}
\end{equation}
\begin{equation}
\mathrm{from ~which}~q_{p} = \frac{9}{7} + \frac{8}{7} q_{c} .
\end{equation}

Thus, we could write:
\begin{equation}
\sqrt{r} \frac{\partial}{\partial r} \left( \nu \Sigma \sqrt{r}\right) \propto (q_{t}+q_{m}+2) r^{q_{t}+q_{m}+3/2}
\end{equation}

From Equation \ref{lb74}, we can derive two different steady states:
\begin{eqnarray}
\label{steady1}q_{t}+q_{m}+2 &=& 0 ~\mathrm{, ~corresponding~to~a~null~flux,}\\
\label{steady2}q_{t}+q_{m}+3/2 &=& 0~\mathrm{, ~corresponding~to~a~uniform~and~non-null~flux.}
\end{eqnarray}

Whereas the expression of the mid-plane temperature for a viscous disk would provide a coupling between $q_{t}$ and $q_{m}$, these two parameters are not related in the case of an irradiated evolving disk. Converging toward a steady state will add a correlation between these two parameters. In the viscous disk case, this will result in setting these two quantities and therefore imposing a temperature and a surface mass density profile. For a simply irradiated disk on the other hand, asymptotic solutions will verify Equations \ref{steady1} or \ref{steady2}.

\bibliography{bibliography}

\begin{thebibliography}{43}
\expandafter\ifx\csname natexlab\endcsname\relax\def\natexlab#1{#1}\fi

\bibitem[{{Alexander} \& {Armitage}(2007)}]{alexander07}
{Alexander}, R.~D. \& {Armitage}, P.~J. 2007, \mnras, 375, 500

\bibitem[{{Alexander} \& {Armitage}(2009)}]{alexander09}
---. 2009, \apj, 704, 989

\bibitem[{{Andrews} {et~al.}(2009){Andrews}, {Wilner}, {Hughes}, {Qi}, \&
  {Dullemond}}]{andrews09}
{Andrews}, S.~M., {Wilner}, D.~J., {Hughes}, A.~M., {Qi}, C., \& {Dullemond},
  C.~P. 2009, \apj, 700, 1502

\bibitem[{{Andrews} {et~al.}(2010){Andrews}, {Wilner}, {Hughes}, {Qi}, \&
  {Dullemond}}]{andrews10}
---. 2010, \apj, 723, 1241

\bibitem[{{Birnstiel} {et~al.}(2010){Birnstiel}, {Dullemond}, \&
  {Brauer}}]{birnstiel10}
{Birnstiel}, T., {Dullemond}, C.~P., \& {Brauer}, F. 2010, \aap, 513, A79

\bibitem[{{Bitsch} {et~al.}(2012){Bitsch}, {Crida}, {Morbidelli}, {Kley}, \&
  {Dobbs-Dixon}}]{bitsch12}
{Bitsch}, B., {Crida}, A., {Morbidelli}, A., {Kley}, W., \& {Dobbs-Dixon}, I.
  2012, ArXiv e-prints

\bibitem[{{Brauer} {et~al.}(2008){Brauer}, {Dullemond}, \&
  {Henning}}]{brauer08}
{Brauer}, F., {Dullemond}, C.~P., \& {Henning}, T. 2008, \aap, 480, 859

\bibitem[{{Calvet} {et~al.}(1991){Calvet}, {Patino}, {Magris}, \&
  {D'Alessio}}]{calvet91}
{Calvet}, N., {Patino}, A., {Magris}, G.~C., \& {D'Alessio}, P. 1991, \apj,
  380, 617

\bibitem[{{Charnoz} \& {Taillifet}(2012)}]{charnoz12}
{Charnoz}, S. \& {Taillifet}, E. 2012, \apj, 753, 119

\bibitem[{{Chiang} \& {Goldreich}(1997)}]{chiang97}
{Chiang}, E.~I. \& {Goldreich}, P. 1997, \apj, 490, 368

\bibitem[{{Ciesla}(2009)}]{ciesla09}
{Ciesla}, F.~J. 2009, \icarus, 200, 655

\bibitem[{{D'Alessio} {et~al.}(2001){D'Alessio}, {Calvet}, \&
  {Hartmann}}]{dalessio01}
{D'Alessio}, P., {Calvet}, N., \& {Hartmann}, L. 2001, \apj, 553, 321

\bibitem[{{D'Alessio} {et~al.}(1998){D'Alessio}, {Canto}, {Calvet}, \&
  {Lizano}}]{dalessio98}
{D'Alessio}, P., {Canto}, J., {Calvet}, N., \& {Lizano}, S. 1998, \apj, 500,
  411

\bibitem[{{Dullemond} {et~al.}(2001){Dullemond}, {Dominik}, \&
  {Natta}}]{dullemond01}
{Dullemond}, C.~P., {Dominik}, C., \& {Natta}, A. 2001, \apj, 560, 957

\bibitem[{{Font} {et~al.}(2004){Font}, {McCarthy}, {Johnstone}, \&
  {Ballantyne}}]{font04}
{Font}, A.~S., {McCarthy}, I.~G., {Johnstone}, D., \& {Ballantyne}, D.~R. 2004,
  \apj, 607, 890

\bibitem[{{Fromang} \& {Papaloizou}(2006)}]{fromang06}
{Fromang}, S. \& {Papaloizou}, J. 2006, \aap, 452, 751

\bibitem[{{Garaud} \& {Lin}(2007)}]{garaud07}
{Garaud}, P. \& {Lin}, D.~N.~C. 2007, \apj, 654, 606

\bibitem[{{Gullbring} {et~al.}(1998){Gullbring}, {Hartmann}, {Briceno}, \&
  {Calvet}}]{gullbring98}
{Gullbring}, E., {Hartmann}, L., {Briceno}, C., \& {Calvet}, N. 1998, \apj,
  492, 323

\bibitem[{{Hartmann} {et~al.}(1998){Hartmann}, {Calvet}, {Gullbring}, \&
  {D'Alessio}}]{hartmann98}
{Hartmann}, L., {Calvet}, N., {Gullbring}, E., \& {D'Alessio}, P. 1998, \apj,
  495, 385

\bibitem[{{Hasegawa} \& {Pudritz}(2011)}]{hasegawa112}
{Hasegawa}, Y. \& {Pudritz}, R.~E. 2011, \mnras, 417, 1236

\bibitem[{{Hayashi}(1981)}]{hayashi81}
{Hayashi}, C. 1981, Progress of Theoretical Physics Supplement, 70, 35

\bibitem[{{Hayashi} {et~al.}(1985){Hayashi}, {Nakazawa}, \&
  {Nakagawa}}]{hayashi85}
{Hayashi}, C., {Nakazawa}, K., \& {Nakagawa}, Y. 1985, in Protostars and
  Planets II, ed. D.~C. {Black} \& M.~S. {Matthews}, 1100--1153

\bibitem[{{Hueso} \& {Guillot}(2005)}]{hueso05}
{Hueso}, R. \& {Guillot}, T. 2005, \aap, 442, 703

\bibitem[{{Hughes} \& {Armitage}(2010)}]{hughes10}
{Hughes}, A.~L.~H. \& {Armitage}, P.~J. 2010, \apj, 719, 1633

\bibitem[{{Isella} {et~al.}(2009){Isella}, {Carpenter}, \&
  {Sargent}}]{isella09}
{Isella}, A., {Carpenter}, J.~M., \& {Sargent}, A.~I. 2009, \apj, 701, 260

\bibitem[{{Jang-Condell}(2008)}]{jc08}
{Jang-Condell}, H. 2008, \apj, 679, 797

\bibitem[{{Jang-Condell} \& {Sasselov}(2004)}]{jc04}
{Jang-Condell}, H. \& {Sasselov}, D.~D. 2004, \apj, 608, 497

\bibitem[{{Kenyon} \& {Hartmann}(1987)}]{kenyon87}
{Kenyon}, S.~J. \& {Hartmann}, L. 1987, \apj, 323, 714

\bibitem[{{Lagage} {et~al.}(2006){Lagage}, {Doucet}, {Pantin}, {Habart},
  {Duch{\^e}ne}, {M{\'e}nard}, {Pinte}, {Charnoz}, \& {Pel}}]{lagage06}
{Lagage}, P.-O., {Doucet}, C., {Pantin}, E., {Habart}, E., {Duch{\^e}ne}, G.,
  {M{\'e}nard}, F., {Pinte}, C., {Charnoz}, S., \& {Pel}, J.-W. 2006, Science,
  314, 621

\bibitem[{{Lynden-Bell} \& {Pringle}(1974)}]{lyndenbellpringle74}
{Lynden-Bell}, D. \& {Pringle}, J.~E. 1974, \mnras, 168, 603

\bibitem[{{Mundy} {et~al.}(2000){Mundy}, {Looney}, \& {Welch}}]{mundy00}
{Mundy}, L.~G., {Looney}, L.~W., \& {Welch}, W.~J. 2000, Protostars and Planets
  IV, 355

\bibitem[{{Owen} {et~al.}(2010){Owen}, {Ercolano}, {Clarke}, \&
  {Alexander}}]{owen10}
{Owen}, J.~E., {Ercolano}, B., {Clarke}, C.~J., \& {Alexander}, R.~D. 2010,
  \mnras, 401, 1415

\bibitem[{{Piau} {et~al.}(2011){Piau}, {Kervella}, {Dib}, \&
  {Hauschildt}}]{piau11}
{Piau}, L., {Kervella}, P., {Dib}, S., \& {Hauschildt}, P. 2011, \aap, 526,
  A100

\bibitem[{{Ruden} \& {Pollack}(1991)}]{ruden91}
{Ruden}, S.~P. \& {Pollack}, J.~B. 1991, \apj, 375, 740

\bibitem[{{Shakura} \& {Sunyaev}(1973)}]{shakura73}
{Shakura}, N.~I. \& {Sunyaev}, R.~A. 1973, \aap, 24, 337

\bibitem[{{Siebenmorgen} \& {Heymann}(2012)}]{siebenmorgen12}
{Siebenmorgen}, R. \& {Heymann}, F. 2012, \aap, 539, A20

\bibitem[{{Tscharnuter} \& {Gail}(2007)}]{tscharnuter07}
{Tscharnuter}, W.~M. \& {Gail}, H.-P. 2007, \aap, 463, 369

\bibitem[{{Vorobyov} \& {Basu}(2007)}]{vorobyov07}
{Vorobyov}, E.~I. \& {Basu}, S. 2007, \mnras, 381, 1009

\bibitem[{{Watanabe} \& {Lin}(2008)}]{watanabe08}
{Watanabe}, S.-i. \& {Lin}, D.~N.~C. 2008, \apj, 672, 1183

\bibitem[{{Weidenschilling}(1977)}]{weiden77}
{Weidenschilling}, S.~J. 1977, \apss, 51, 153

\bibitem[{{Yang} \& {Ciesla}(2012)}]{yang12}
{Yang}, L. \& {Ciesla}, F.~J. 2012, Meteoritics and Planetary Science, 47, 99

\bibitem[{{Zhu} {et~al.}(2008){Zhu}, {Hartmann}, {Calvet}, {Hernandez},
  {Tannirkulam}, \& {D'Alessio}}]{zhu08}
{Zhu}, Z., {Hartmann}, L., {Calvet}, N., {Hernandez}, J., {Tannirkulam}, A.-K.,
  \& {D'Alessio}, P. 2008, \apj, 684, 1281

\bibitem[{{Zhu} {et~al.}(2010){Zhu}, {Hartmann}, {Gammie}, {Book}, {Simon}, \&
  {Engelhard}}]{zhu101}
{Zhu}, Z., {Hartmann}, L., {Gammie}, C.~F., {Book}, L.~G., {Simon}, J.~B., \&
  {Engelhard}, E. 2010, \apj, 713, 1134

\end{thebibliography}
\end{document}